\documentclass{aa}  
\usepackage{graphicx}
\usepackage{xcolor}
\usepackage{mathrsfs}
\usepackage{longtable}
\usepackage{txfonts}
\newcommand{\hthp}{\ensuremath{\rm H_3^+}\xspace}
\newcommand{\ammo}{\ensuremath{\rm NH_3}\xspace}
\newcommand{\nthp}{\ensuremath{\rm N_2H^+}\xspace}
\newcommand{\ntdp}{\ensuremath{\rm N_2D^+}\xspace}
\newcommand{\dcop}{\ensuremath{\rm DCO^+}\xspace}
\newcommand{\form}{\ensuremath{\rm H_2CO}\xspace}
\newcommand{\thioform}{\ensuremath{\rm H_2CS}\xspace}
\newcommand{\cycloprop}{\ensuremath{\rm C_3H_2}\xspace}
\newcommand{\meth}{\ensuremath{\rm CH_3OH}\xspace}

\newcommand{\percc}{\ensuremath{\rm cm^{-3}}\xspace}

\def\htwo{\rm H_2}
\def\persqcm{\rm cm^{-2}}
\def\ms{\rm m\,s^{-1}}

\def\diaz{\rm N_2H^+}
\def\ddiaz{\rm N_2D^+}
\def\htwo{\rm H_2}
\def\dcoplus{\rm DCO^+}
\def\ndthree{\rm ND_3}

\begin{document} 

   \title{Hunting pre-stellar cores with APEX: overview}

\author{P. Caselli\inst{1} \and S. Spezzano\inst{1}  \and E. Redaelli\inst{2,1} \and J. Harju\inst{1,3} \and D. Arzoumanian\inst{4} \and F. Lique\inst{5} \and O. Sipil\"a\inst{1} \and J.~E. Pineda\inst{1} \and E. Wirstr\"om\inst{6} \and F. Wyrowski\inst{7} \and A. Belloche\inst{7}}  

\institute{Max-Planck-Institut f\"ur Extraterrestrische Physik, Giessenbachstrasse 1, 85748 Garching, Germany \and 
European Southern Observatory, Karl-Schwarzschild-Strasse 2, 85748 Garching, Germany \and 
Department of Physics, P.O. Box 64, FI-00014, University of Helsinki, Finland \and National Astronomical Observatory of Japan, Osawa 2-21-1, Mitaka, Tokyo 181-8588, Japan \and Univ. Rennes, CNRS, IPR (Institut de Physique de Rennes) – UMR 6251, 35000 Rennes, France \and Department of Space, Earth, and Environment, Chalmers University of Technology, 412 96 Gothenburg, Sweden, \and Max-Planck-Institut f\"ur Radioastronomie, Auf dem H\"ugel 69, 53121 Bonn, Germany}

\abstract 
{\textcolor{black}{Pre-stellar cores are centrally concentrated starless cores on the verge of star formation and they represent the  initial conditions for star and planet formation. Pre-stellar cores host an active organic chemistry and isotopic fractionation, kept stored into thick icy mantles, which can be inherited by the future protoplanetary disks and planetesimals. It is therefore important to study pre-stellar cores, but this is difficult as they are short-lived and thus rare. So far, only a few pre-stellar cores have been studied in detail, with special attention being paid to the prototypical pre-stellar core L1544 in the Taurus Molecular Cloud.}}
{\textcolor{black}{The aim is to identify nearby ($<$200\,pc) pre-stellar cores in an unbiased way, to build a sample that can then be studied in detail.  This will also allow us to explore the effect of the environment on the chemical and physical structure of pre-stellar cores.}}
{\textcolor{black}{We first used the {\em Herschel} Gould Belt Survey archival data, selecting all those starless cores with central H$_2$ number densities higher than or equal to 3$\times$10$^5$\,\percc, the density of L1544 within the {\em Herschel}} beam of 20\arcsec. The selected 40 (out of 1746) cores have then been observed in \nthp(3-2) and \ntdp(4-3) using the APEX antenna.}
{\textcolor{black} Following a simple analysis, a total of 17 {\it bona-fide} (i.e., with a deuterium fraction larger than 10\%) pre-stellar cores have been identified. Other 16 objects can also be considered pre-stellar, as they are dynamically evolved starless cores, but their deuterium fractions is relatively low ($<$10\%); thus, they deserve further scrutiny to unveil the source of the low deuteration. Of the remaining 7 objects, six have been found associated with a young stellar object, and one (CrA\,151) presents hints of a very young (or very low luminosity) stellar object.} 
{\textcolor{black}{Dust continuum emission, together with spectroscopic observations of \nthp(3-2) and \ntdp(4-3), is a powerful tool to identify pre-stellar cores in molecular clouds. Detailed modeling of the physical structure of the objects is now required for reconstructing the chemical composition as a function of radius. This work has provided a statistically significant sample of 33 pre-stellar cores, a crucial step in the understanding of the process of star and planet formation.}}

\titlerunning{}
\maketitle

%

\section{Introduction} \label{Sec:Introduction}

Stellar systems are the product of the collapse of dense cores within interstellar molecular clouds \citep[e.g.,][]{BM1989,BT2007}. It is therefore important to study dense cores, as the initial conditions for star and planet formation are to be found there, including all the ingredients participating in the spectacular chemical and physical evolution from interstellar clouds to planets such as our Earth. Particularly interesting are starless cores with centrally concentrated density profiles and central densities higher than 10$^5$ \percc, which can become unstable against gravitational collapse \citep{KetoCaselli2008}. These cores are called "pre-stellar" as they are expected to contract and proceed toward star formation, while the less dense ones could be thermally supported and pulsating \citep[such as B68;][]{Alves2001,Lada2003}, pressure-confined and unbound \citep[such as the starless cores in the Pipe Nebula;][]{Lada2008}, or transient structures with evidence of expansion motions \citep[as in the case of L1517B;][]{Tafalla2004} or simply at an earlier stage of evolution \citep{TafallaSantiago2004}\footnote{We note that our definition of pre-stellar core is more restrictive than the one given by \cite{Andre2000}, as they consider pre-stellar all dense cores which are gravitationally bound.}.

Pre-stellar cores have clear temperature gradients, from outer edge values around 10-13\,K and central values around 6-8\,K \citep[e.g.,][]{Crapsi2007,Pagani2007,Launhardt2013}, well understood by radiative transfer models of externally illuminated Bonnor-Ebert \citep[BE;][]{Bonnor1956,Ebert1955} spheres \citep[e.g.,][]{Evans2001,Zucconi2001,Goncalves2004}. The low temperatures and high densities favor the freeze-out of molecules, which becomes "catastrophic" in pre-stellar cores within the central few thousand astronomical units \citep{Caselli1999,Caselli2022,Bacmann2002,Redaelli2019,Pineda2022,Lin2023a}. This is important, as it implies that just before star formation dust grains are enshrouded in thick icy mantles, where volatiles crucial for mineral evolution and the synthesis of pre-biotic material (in particular water and organic molecules) are then delivered to the future protostellar disk. Indeed, large amounts of water \citep{Caselli2012} and organics \citep[e.g.,][]{Oberg2010,Bacmann2012,Cernicharo2021,JimenezSerra2016,JimenezSerra2021,McGuire2018,ScibelliShirley2020,Punanova2022,Megias2023,Scibelli2024} are known to be present in these and similar environments, as recently confirmed by the {\em{James Webb Space Telescope}} (JWST) sharp view of ice components along the line of sight of stars behind the Chamaeleon I dark molecular cloud \citep{McClure2023}. 

A large amount of molecular freeze-out and low temperatures within pre-stellar cores also boost deuterium fractionation \citep{DalgarnoLepp1984,Roberts2003,Walmsley2004,Sipila2010,Aikawa2012,Taquet2012}, leading to deuterium fractions orders of magnitude larger than the cosmic D/H value \citep[$\simeq$1.5$\times$10$^{-5}$;][]{Moos2002}.  Thus, copious amounts of deuterium atoms and deuterated species such as \dcop, \ntdp, DCN, DNC as well as singly and multiply deuterated forms of \hthp, \ammo, \form, \thioform, c-\cycloprop and \meth are produced \citep[e.g.,][]{Caselli2002,Caselli2003,Lis2002,Bacmann2003,Vastel2004,Crapsi2005,Parise2011,Spezzano2013,Chantzos2018,Harju2017,Bizzocchi2014,Redaelli2019,Ambrose2021,Spezzano2022,Giers2022,Giers2023,Lin2023b}. Large deuterium fractions of simple species are also measured toward protostellar objects, especially toward the young Class 0 sources still surrounded by the cold and dense envelope, part of the original pre-stellar core. For example, \cite{Emprechtinger2009} found \nthp D-fractions $\geq$15\% for the youngest objects and lower values ($\simeq$3\%) at the Class 0-I borderline \citep[see also][]{Friesen2010,Friesen2013,Punanova2016,Chantzos2018,Giers2023,Mercimek2025}. Closer to the protostar, within the so-called "hot corino", where the dust temperature increases above the evaporation temperature of water \citep[$\sim$100\,K; e.g.,][]{Ceccarelli2007}, complex organic molecules, and in particular methanol, also present significant D-fractions \citep[e.g.,][]{Parise2002,Parise2004,Manigand2020,FerrerAsensio2023,Bunn2025}. This can be explained if the evaporated ices are those formed during the pre-stellar phase, where the high atomic D/H ratio \citep[due to the dissociative recombination of the deuterated forms of \hthp,][]{Roberts2003,Walmsley2004} allows efficient surface deuteration of surface species, such as CO, producing deuterated organics, such as CH$_3$OH \citep{Tielens1983,CaselliStancheva2002,Taquet2012,Riedel2023}. Lower D-fractions of organics are measured at later stages of protostellar evolution \citep{Bianchi2017}, possibly due to gas-phase processes at high temperature, which tend to decrease the deuteration. That pre-stellar core ices are at least partially maintained through the whole process of star and planet formation is also demonstrated by the need to invoke pre-stellar chemistry to explain the water D/H ratio in our oceans \citep[about ten times higher than the cosmic D/H;][]{Cleeves2014} and the large methanol D/H ratio measured in comet 67P/Churyumov-Gerasimenko \citep{Drozdovskaya2021}. 

Despite the important role played by pre-stellar cores in sculpting the physical structure and chemical composition of stellar systems, only a few have been found and investigated with single dish telescopes \citep[e.g.,][]{Crapsi2005,Pagani2005,Spezzano2020,Lin2023a,Lin2023b} and only one studied extensively, including interferometric observations: L1544 in the Taurus Molecular Cloud \citep[e.g.,][]{Caselli1999,Caselli2002,Caselli2019,Caselli2022,Ohashi1999,WilliamsMyers1999,Spezzano2016,Spezzano2017,Spezzano2022,Redaelli2019,Redaelli2021,Redaelli2022}. The reason for this is that pre-stellar cores are rare because their life expectancy is short, i.e. the passage from the pre- to proto-stellar phase is fast \citep[between 1 to 10 free-fall times;][]{Andre2014}. For example, the life expectancy of L1544 is estimated to be less than 50,000\,yr, based on the free-fall time relative to the central structure, called the "kernel", revealed by the Atacama Large Millimeter and sub-millimeter Array \citep[ALMA;][]{Caselli2019} and evolutionary models of magnetized cores \citep[e.g.,][]{Tassis2007}. Therefore, we decided to start a new hunt for pre-stellar cores in nearby molecular clouds, using the unbiased view of the {\em{Herschel Space Observatory}}, as provided by the Herschel Gould Belt Survey \citep[HGBS;][]{Andre2010} legacy program, and our detailed knowledge of the prototypical pre-stellar core L1544. With this new unbiased method, we identified 40 pre-stellar core candidates (Sect.\,\ref{Subsec:sample}), which have then been observed in high-excitation \nthp and \ntdp lines with the Atacama Pathfinder EXperiment (APEX) to assess their pre-stellar nature (Sect.\,\ref{Subsec:apex}). In Section \ref{Sec:results} we will present the data, show spectra, provide a preliminary estimate of the D/H ratio and physical conditions of the gas traced by the selected lines and finally identify the {\it bona-fide} pre-stellar cores. A discussion will follow in Sect.\,\ref{Sec:discussion} and conclusions in Sect.\,\ref{Sec:conclusion}. The aim of this paper is to present the project overview, introduce the sample and show the APEX data. Detailed studies of the physical and chemical structure of individual pre-stellar cores are presented in \cite{Redaelli2025}, \cite{Spezzano2025}, and future papers. 

\section{Observations}

\subsection{The Sample Selection} \label{Subsec:sample}


As mentioned in the introduction, our definition of pre-stellar cores is more restrictive than the one given by \cite{Andre2000}, as we consider "pre-stellar" all dense starless cores which are gravitationally bound \citep[as in ][]{Andre2000} {\it and} have centrally concentrated density profiles with central densities higher than 10$^5$\,\percc, thus thermally supercritical \citep{KetoCaselli2008} and likely contracting, as in the case of the prototypical pre-stellar core L1544 \citep{Keto2015}. 

Our initial sample included all nearby (distance $<$200\,pc) starless cores identified in the Herschel Gould Belt Survey Archive \citep[HGBSA;][and the publicly available catalogs
http://gouldbelt-herschel.cea.fr/archives]{Andre2010}. The total number of starless cores (including starless, prestellar and prestelllar candidates source in the HGBS catalogs) is 1746 in the Ophiuchus, Corona Australis, Taurus, and Lupus Molecular Clouds. These catalogs provide the identified core properties including their number densities \citep[see e.g.,][]{Konyves2015}. The method to derive the core number density is the following: First, the peak column density ($N$) is derived from graybody spectral energy distribution (SED) fits to the peak flux densities in a common beam of 36\arcsec\, at the {\em Herschel} four wavelengths (160, 250, 350, 500\,$\mu$m). The beam-averaged peak number density ($n$) is derived from the peak column density ($N$) and the radius ($R$) of the cores assuming a Gaussian spherical distribution $n$=$\sqrt{4ln2/\pi}\times N/R$, where $R$ is the core radius measured at the 20\arcsec\, angular resolution. We then computed the number density of L1544 within the central 20\arcsec\, (the {{\em Herschel}} resolution), and found  $n$(H$_2$)$_{20\arcsec}^{\rm L1544}$=3$\times10^5$\,\percc. The next step was to extract from the HGBSA sample those objects with $n$(H$_2$)$_{20\arcsec}$ $\ge$ $n$(H$_2$)$_{20\arcsec}^{\rm L1544}$. Only 44 cores fulfilled this criterium. After excluding four cores having protostars in their immediate vicinity, the final sample consists of 40 pre-stellar core candidates. 

Table \ref{Tab:targets} presents the whole sample of 40 pre-stellar core candidates, selected for observations with the Atacama Pathfinder EXperiment (APEX; see Sect.\,\ref{Subsec:apex}): the HGBSA name is in column 1, J2000 equatorial coordinates are in columns 2 and 3, column 4 lists the H$_2$ column density derived from the {{\em Herschel}} data at the resolution of the 500\,$\mu$m surface brightness maps (36\arcsec.3), the observing setup is specified in column 5 (see notes for the frequency ranges and receivers used), column 6 shows other names used in the literature, with corresponding references reported in the notes to the table. We immediately note that 6 of the 40 pre-stellar core candidates (marked by an asterisk in Table \ref{Tab:targets}) are associated with known very young stellar objects (YSOs). In addition, one of the cores (CrA\,044) has an unidentified mid-infrared source (detected at 8 and 24\,$\mu$m with {\em Spitzer}) within its radius. However, higher angular resolution spectroscopic observations are needed to confirm if this source is embedded in the CrA\,044 core and we consider CrA\,044 starless in this study. The core properties derived from the HGBS data are listed in Appendix\,\ref{Appendix}. There, we also give criteria for association with a YSO.

\begin{table*} 
{\centering
\caption[]{Pre-stellar core candidates observed with APEX.}
\label{Tab:targets}
  \begin{tabular}{lrrrll} \hline
   \noalign{\smallskip}

HGBSA & \multicolumn{1}{c}{R.A.} & \multicolumn{1}{c}{Dec.} & \multicolumn{1}{c}{$N(\htwo)^a$} & setup$^b$ & other\\ 
name$^c$     & \multicolumn{2}{c}{(J2000.0)} & \multicolumn{1}{c}{($\times 10^{21}\,\persqcm$)} &  & designations\\ 
\hline 
\noalign{\smallskip}

Tau\,109 & 4 14 02.98 &  28 09 47.0 &  40.1 &S  &J041403.0+280947$^{\rm HGBS,M}$ \\  
Tau\,410 & 4 18 32.99 &  28 28 29.0 &  33.6 &S,F&J041833.0+282829$^{\rm HGBS,M}$\\ 
Tau\,420 & 4 18 40.32 &  28 23 16.0 &  50.7 &S,F&J041840.3+282316$^{\rm HGBS,M}$, J041840.1+282324$^{\rm E}$ \\ 
\noalign{\smallskip}
Lup\,288 &15 44 59.03 & -34 17 05.4 & 104.4 &S  &Lup1 C2$^{\rm G}$\\ 
Lup\,039 &16 01 54.81 & -41 52 46.1 &  41.6 &S  &\\
Lup\,032 &16 09 17.49 & -39 07 38.1 &  39.8 &S  &\\ 
\noalign{\smallskip}
Oph\,082 &16 26 24.06 & -24 21 54.6 &  50.9 &S  & J162624.0-242154$^{\rm HGBS,L}$, A-MM4$^{\rm M}$ \\   
Oph\,087 &16 26 26.33 & -24 22 25.5 &  51.8 &S  & J162626.3-242225$^{\rm HGBS,L}$, A-N2$^{\rm D04}$\\ 
Oph\,091$^*$ &16 26 27.65 & -24 23 59.3 & 378.8 &S,F& Oph\,A-SM1$^{\rm A}$\\  
Oph\,129 &16 26 43.71 & -24 17 24.8 &  37.6 &S,F& A-MM18$^{\rm M}$ \\    
Oph\,146 &16 26 48.63 & -24 29 36.9 &  41.3 &S  & J162648.6-242936$^{\rm HGBS,L}$,C-MM8$^{\rm M}$  \\  
Oph\,169$^*$ &16 26 58.91 & -24 34 22.3 &  77.9 &S  & C-A1$^{\rm F09}$ \\  
Oph\,178 &16 27 05.19 & -24 39 17.6 &  53.3 &S,F& J162705.1-243917$^{\rm HGBS,L}$, L1688-d5$^{\rm C}$ \\  
Oph\,196 &16 27 12.56 & -24 29 49.0 &  70.4 &S,F& J162712.5-242949$^{\rm HGBS,L}$,B1-A2$^{\rm F09}$ \\  
Oph\,201 &16 27 14.51 & -24 30 26.1 &  59.2 &S  & J162714.5-243026$^{\rm HGBS,L}$,J162714.6-243020$^{\rm D08}$ \\
Oph\,215 &16 27 20.13 & -24 27 12.2 &  62.6 &S  & J162720.1-242712$^{\rm HGBS,L}$,L1688 58$^{\rm Ke}$ \\ 
Oph\,219 &16 27 21.56 & -24 39 49.1 &  25.5 &S  & J162721.5-243949$^{\rm HGBS,L}$,F-MM1$^{\rm M}$ \\ 
Oph\,227 &16 27 24.99 & -24 27 08.5 &  59.6 &S,F& J162724.9-242708$^{\rm HGBS,L}$,B2-MM6a$^{\rm Ka}$, B2-N2$^{\rm F10}$\\ 
Oph\,229 &16 27 25.39 & -24 26 54.6 &  44.9 &S,F& J162725.3-242654$^{\rm HGBS,L}$,B2-MM6b$^{\rm Ka}$, B2-A5$^{\rm F09}$\\ 
Oph\,237$^*$ &16 27 28.18 & -24 26 36.0 & 105.9 &S,F& J162728.1-242636$^{\rm HGBS,L}$, B2-N4$^{\rm F10}$, B2-MM9b$^{\rm Ka}$\\ 
Oph\,238$^*$ &16 27 28.19 & -24 27 12.5 &  67.6 &S,F& J162728.1-242712$^{\rm HGBS,L}$, B2-MM8b$^{\rm Ka}$\\ 
Oph\,246 &16 27 32.93 & -24 26 30.7 & 160.7 &S,F& J162732.9-242630$^{\rm HGBS,L}$,B2-MM14a$^{\rm Ka}$ \\ 
Oph\,316 &16 28 29.08 & -24 19 16.7 &  67.6 &S,F& J162829.0-241916$^{\rm HGBS,L}$,D-MM1$^{\rm M}$, MMS047$^{\rm S}$ \\ 
Oph\,319 &16 28 31.64 & -24 18 06.7 &  54.8 &S  & J162831.6-24180$^{\rm HGBS,L}$,D-MM4$^{\rm M}$\\ 
Oph\,332 &16 28 57.96 & -24 20 53.6 &  50.3 &S  & J162857.9-242053$^{\rm HGBS,L}$\\ 
Oph\,385 &16 31 38.28 & -24 49 49.2 &  79.4 &S,F& J163138.2-244949$^{\rm HGBS,L}$,J163137.7-244947$^{\rm P}$\\ 
Oph\,387 &16 31 38.83 & -24 50 09.0 &  35.6 &S,F& J163138.8-245009$^{\rm HGBS,L}$\\ 
Oph\,410 &16 31 57.14 & -24 57 14.8 &  34.8 &S  & J163157.1-245714$^{\rm HGBS,L}$\\ 
Oph\,412 &16 31 57.76 & -24 57 53.6 &  66.6 &S  & J163157.7-245753$^{\rm HGBS,L}$\\ 
Oph\,455 &16 32 21.66 & -24 27 43.0 & 106.6 &S  & J163221.6-242743$^{\rm HGBS,L}$,J163221.6-242739$^{\rm P}$\\ 
Oph\,464 &16 32 29.07 & -24 29 09.0 & 228.2 &S,F& IRAS16293E$^{\rm Ki, HGBS,L}$\\ 
Oph\,485 &16 32 46.70 & -23 52 27.4 &  50.0 &S  & J163246.7-235227$^{\rm HGBS,L}$\\ 
\noalign{\smallskip}
CrA\,021 &19 01 10.83 & -36 54 15.3 &  32.8 &S,F& J190110.8-365415$^{\rm HGBS,B }$ \\    
CrA\,038 &19 01 46.10 & -36 55 35.7 &  37.3 &S,F&J190146.1-365535$^{\rm HGBS,B}$, SMM6$^{\rm N}$ \\  
CrA\,040 &19 01 47.28 & -36 56 39.8 & 109.5 &S,F&J190147.2-365639$^{\rm HGBS,B}$ \\  
CrA\,044 &19 01 54.45 & -36 57 48.9 &  48.1 &S,F& J190154.4-365748$^{\rm HGBS,B}$, SMM1A$^{\rm N, Gro}$\\  
CrA\,047$^*$ &19 01 55.86 & -36 57 46.9 &  73.0 &S,F& J190155.8-365746$^{\rm HGBS,B}$ , MMS13$^{\rm Chini}$ \\  
CrA\,050$^*$ &19 01 58.94 & -36 57 09.9 &  42.8 &S,F& SMM2$^{\rm N,H}$ \\  
CrA\,066 &19 02 17.51 & -37 01 35.8 &  39.1 &S,F& J190217.5-370135$^{\rm HGBS,B}$, MMS21$^{\rm Chini}$ \\  
CrA\,151 &19 10 20.17 & -37 08 27.0 &  24.1 &S,F& J191020.1-370826$^{\rm HGBS,B}$,  Core 5$^{\rm Y}$, SL42$^{\rm H-U}$ \\ \hline
\noalign{\smallskip}
\end{tabular}

}

$^a$ The peak $\htwo$ column density at the peak derived from {\sl Herschel} data at the
original resolution of the $500\,\mu$m surface brightness maps
($36\farcs3$). The dust temperature is reported in Table\,\ref{derived_core_properties}.  The values are taken from catalogues provided by the {\sl Herschel} Gould Belt Survey
Archive (http://www.herschel.fr/cea/gouldbelt).

$^b$ {\bf S:} The Sepia345 receiver used in two tunings covering the
frequency ranges 1) 271.8-279.7\,GHz (LSB) and 288.0-295.9\,GHz (USB),
and 2) 287.3-295.3\,GHz (LSB) and 303.6-311.5\,GHz (USB); so the
frequencies $\sim 288-295$\,GHz were observed twice. {\bf F:} The nFLASH230/460 dual channel receiver tuned to cover the frequency ranges 224.4-232.3\,GHz (LSB), 240.6-248.5\,GHz (USB), 462.2-466.2\,GHz (LSB), and 474.6-478.6\,GHz (USB).

$^c$ An asterisk next to the HGBSA name indicates association with a young stellar object (YSO); see main text.

{\bf References:} 
$^{\rm A}$ \cite{1993ApJ...406..122A};
$^{\rm C}$ \cite{2019ApJ...877...93C};
$^{\rm Chini}$ \cite{2003A&A...409..235C};
$^{\rm D04}$ \cite{2004ApJ...617..425D};
$^{\rm D08}$ \cite{2008ApJS..175..277D};
$^{\rm E}$ \cite{2019MNRAS.485.2895E}; 
$^{\rm F09}$ \cite{2009ApJ...697.1457F};
$^{\rm F10}$ \cite{2010ApJ...708.1002F};
$^{\rm G}$ \cite{2015A&A...584A..36G};
$^{\rm Gro}$ \cite{2007ApJ...670..489G};
$^{\rm H}$ \cite{2008A&A...488..987H};
$^{\rm HGBS,B }$ \cite{2018A&A...615A.125B};
$^{\rm HGBS,M}$ \cite{2016MNRAS.459..342M};
$^{\rm HGBS,L}$ \cite{2020A&A...638A..74L};
$^{\rm H-U}$ \cite{2013ApJ...763...45H};
$^{\rm Ka}$ \cite{2019ApJ...871...86K};
$^{\rm Ke}$ \cite{2019ApJ...874..147K};
$^{\rm Ki}$ \cite{2017ApJ...838..114K};
$^{\rm M}$ \cite{1998A&A...336..150M}; 
$^{\rm N}$ \cite{2005MNRAS.357..975N}; 
$^{\rm P}$ \cite{2015MNRAS.450.1094P};
$^{\rm S}$ \cite{2006A&A...447..609S};
$^{\rm Y}$ \cite{1999PASJ...51..911Y}

 \end{table*}

\subsection{APEX observations} \label{Subsec:apex}


APEX observations were carried out during 2022 and 2023 under the projects O-0110.F-9310A-2022, M-0110.F-9501C-2022, and M-0110.F-9501C-2023 (PI: Caselli). In the initial survey, the center positions of all 40 cores of the original HGBS sample were observed using the SEPIA345 receiver \citep{Meledin2022} using frequency setups that cover the $\diaz$\,(3-2) and $\ddiaz$\,(4-3) lines. These were single-point position-switched observations. The OFF positions were selected 5 arcmin away from the target in a direction that appeared free of dense gas judging from {\sl Herschel} far-infrared maps. The tuning range of SEPIA345 is $272-376$\,GHz. The receiver has two 8\,GHz wide IF outputs (upper and lower sidebands) per polarization, separated by 8\,GHz. These were recorded with altogether 8 Fast Fourier Transform Spectrometer (FFTS) units with a bandwidth of 4\,GHz each. The FFTS units were configured so that the adjacent spectrometer bands overlap by approximately 100\,MHz, so the total bandwidth per sideband and polarization is 7.9\,GHz.  Each 4\,GHz spectrometer band was resolved into 65 536 spectral channels of 61.03\,kHz in width. Using two tunings, the approximate frequency ranges covered by SEPIA are $272-280$\,GHz, $288-296$\,GHz, and $304-312$\,GHz (the range $288-296$\,GHz was observed twice). In addition to $\diaz$\,(3-2) and $\ddiaz$\,(4-3), for example the $\dcoplus$\,(4-3), CS\,(6-5), and meta-$\ndthree\,(1_0-0_0)$ lines are included in these frequency ranges.

Subsequent ON-OFF observations towards 23 HGBS targets selected based on their $\diaz$\,(3-2) and $\ddiaz$\,(4-3) spectra were made using the dual channel nFLASH230/460 receiver tuned to measure the $\ddiaz$\,(3-2), $\ddiaz$\,(6-5), and $\diaz$\,(5-4) lines. This setup was also used to observe the archetypical prestellar core L1544 in Taurus (e.g., \citealt{Caselli2019}). In the analysis of this source, we also use the $\diaz$\,(3-2) spectrum observed with the IRAM 30\,m telescope by \cite{Redaelli2019}. The nFLASH230 receiver is tunable between 196 and 281\,GHz, and its IF outputs (LSB and USB) are 8\,GHz wide when two polarizations are measured simultaneously. The tuning range of the nFLASH460 receiver is $378-508$\,GHz, and the IF bandwidth is 4\,GHz. For both nFLASH receivers, the LSB and USB are separated by 8\,GHz. With a single tuning, the nFLASH230/460 receivers covered the spectral ranges $224-232$\,GHz, $241-249$\,GHz, $462-466$\,GHz, and $475-479$\,GHz. In addition to $\ddiaz$\,(3-2), $\ddiaz$\,(6-5), and $\diaz$\,(5-4) lines, these frequency ranges contain several other lines, for example, CN\,($N=2-1$), CS\,(5-4), and C$^{17}$O$\,(2-1)$. 

In this overview paper, we focus the attention on the \nthp and \ntdp lines (see Table \ref{Tab:lines}), as these allow us to follow an analysis similar to the one carried out by \cite{Crapsi2005}, although the higher J transitions considered here selectively trace material at densities higher than 10$^5$\,\percc, possibly including the {\it kernel}, the central few thousand au flattened structure discovered in L1544 with ALMA \citep{Caselli2019,Caselli2022}. All the other molecules and transitions detected in the receiver setups listed in column 6 of Table \ref{Tab:targets} will be presented in the papers focusing on individual sources \citep[see first papers by][]{Redaelli2025,Spezzano2025}. Table \ref{Tab:lines} lists the \nthp and \ntdp lines observed with APEX (columns 1 and 2), the transition frequencies \citep[column 3, from the Cologne Database for Molecular Spectroscopy;][]{Endres2016}, the critical density of the transition ($n_{\rm crit}$, column 4), the receiver used (column 5), the half power beam width (HPBW, column 6), the main beam efficiency (column 7), the velocity resolution (column 8) and the root mean square noise (rms, column 9). The critical densities of the observed transitions of $\diaz$ and $\ddiaz$ shown in Table~2 were estimated at 10\,K, using the optically thin formula of Shirley (\citeyear{2015PASP..127..299S}, Eq.~(4)). This formula takes into account all collisionally induced transitions starting from the upper transition level of the emission line.  The line frequencies were taken from \cite{2009A&A...494..719P}, and the  collisional (de-)excitation rate coefficients were adopted from \cite{2005A&A...432..369S} and \cite{2015MNRAS.446.1245L}.

\begin{table*}
{\centering
\caption[]{Observed molecule, transition, frequency, critical density ($n_{\rm crit}$), half power beam width (HPBW), main beam efficiency ($\eta_{\rm MB}$), velocity resolution ($\Delta\upsilon$), and root mean square noise (rms).}
\label{Tab:lines}
\begin{tabular}{llcclcccc}  \hline
   \noalign{\smallskip}
  Molecule & Transition & Frequency & $n_{\rm crit}$ & Receiver & HPBW & $\eta_{\rm MB}$ & $\Delta\upsilon$ & rms \\
  &  &  (MHz) & ($\percc$) &  & ($\arcsec$) &  & ($\ms$) & (mK) \\ \hline
   \noalign{\smallskip}
  $\nthp$ & $J=3-2$ & 279511.832 & $1.5\times10^6$ & SEPIA345  & 22 & 0.77 & 65 & 40 \\
        & $J=5-4$ & 465824.947 & $5.6\times10^6$ & nFLASH460 & 13 & 0.61 & 39 & 40 \\
          \noalign{\smallskip} 
 $\ntdp$ & $J=3-2$ & 231321.828 & $7.5\times10^5$ & nFLASH230 & 27 & 0.81 & 79 & 15 \\
        & $J=4-3$ & 308422.294 & $1.6\times10^6$ & SEPIA345  & 20 & 0.74 & 59 & 32 \\
        & $J=6-5$ & 462603.932 & $3.8\times10^6$ & nFLASH460 & 13 & 0.61 & 39 & 37 \\ \hline
   \noalign{\smallskip}
\end{tabular}

}

\end{table*}

\section{Results} \label{Sec:results}

\subsection{Spectra} \label{Subsec:spectra}
\begin{figure*} 
\begin{center}
\includegraphics[width=14cm]{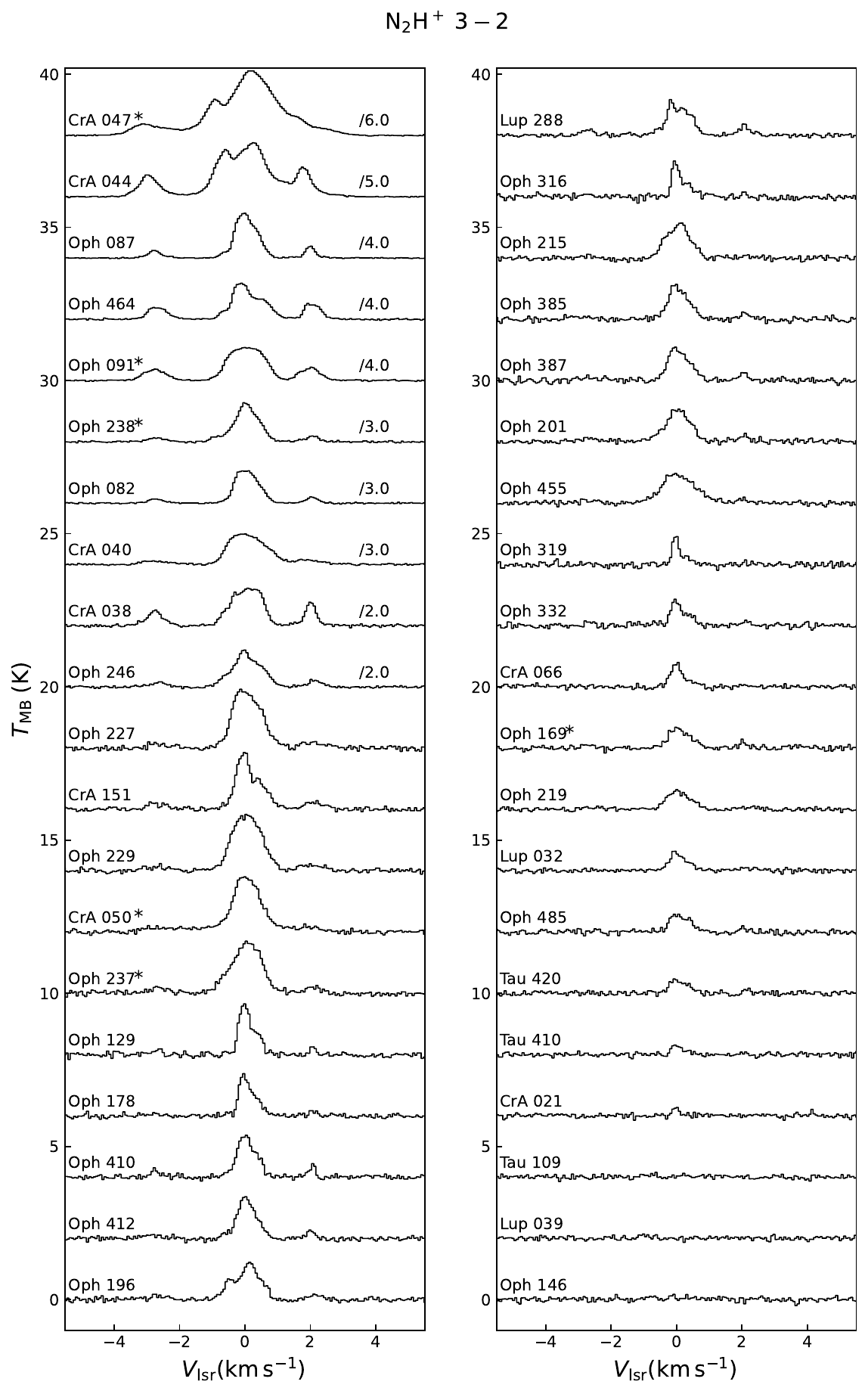} 
\end{center}
\caption{Spectra of \nthp (3-2) toward the whole sample of pre-stellar core candidates in Table \ref{Tab:targets}. From top to bottom and then from left to right, the spectra are ordered in integrated intensity, with CRA\,047 being the strongest. The spectra have also been displaced in intensity by multiples of 2\,K and centered at 0 LSR velocity, to allow comparison. Note that the spectra of the first 10 objects have been divided by factors between 6 and 2 (see labels) for clarity. Asterisks next to the names indicate the association with a young stellar object (YSO).}
\label{Fig:n2hp32}
\end{figure*}

\begin{figure*}
\begin{center}
\includegraphics[width=14cm]{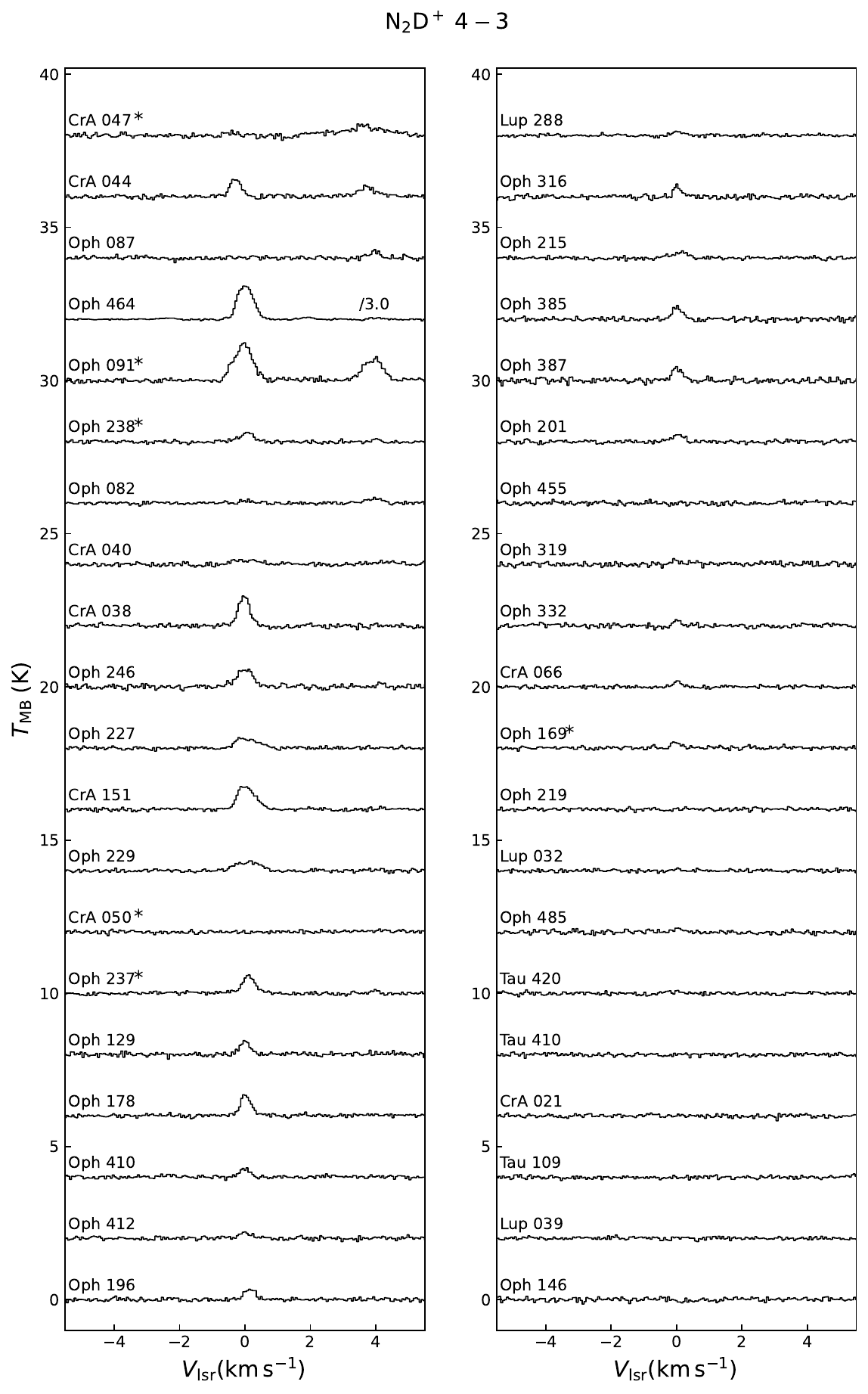} 
\end{center}
\caption{Spectra of \ntdp (4-3) toward the whole sample of pre-stellar core candidates in Table \ref{Tab:targets}. The spectra follow the same order as in Fig.\,\ref{Fig:n2hp32}. The spectra have also been displaced in intensity by multiples of 2\,K and centered at 0 LSR velocity, to allow comparison. Note that the spectrum of Oph\,464 has been divided by a factor of 3 (see label) for clarity.}
\label{Fig:n2dp43}
\end{figure*}

The whole sample of pre-stellar core candidates in Table \ref{Tab:targets} has been observed in \nthp (3-2) and \ntdp(4-3), while a selection of 23 objects has been observed in \nthp (5-4), \ntdp (3-2) and (6-5). We focus here on the \nthp (3-2) and \ntdp (4-3) lines, as they have very similar critical densities (larger than the threshold of 3$\times$10$^5$\,\percc; see Sect. \ref{Subsec:sample}) and they can be observed with similar angular resolution (see Table \ref{Tab:lines}), thus allowing a more accurate determination of the deuterium fraction, assuming that they are tracing the same gas (see Section \ref{Sec:discussion} for more details on this). The spectra of \nthp (3-2) and \ntdp (4-3) towards the whole sample are shown in Figures \ref{Fig:n2hp32} and \ref{Fig:n2dp43}, respectively. They are presented in order of decreasing \nthp (3-2) line intensity. Note that the \nthp (3-2) line is detected almost everywhere (37 out of 40 objects, i.e., in 93\% of the sample). \ntdp (4-3) is detected in 27 objects (68\% of the sample) and the intensity does not follow the \nthp (3-2) trend, suggesting significant variations in the deuterium fraction among sources (see Sect.\,\ref{Subsec:CTex}). 

\begin{figure*}
\begin{center}
\includegraphics[width=15cm]{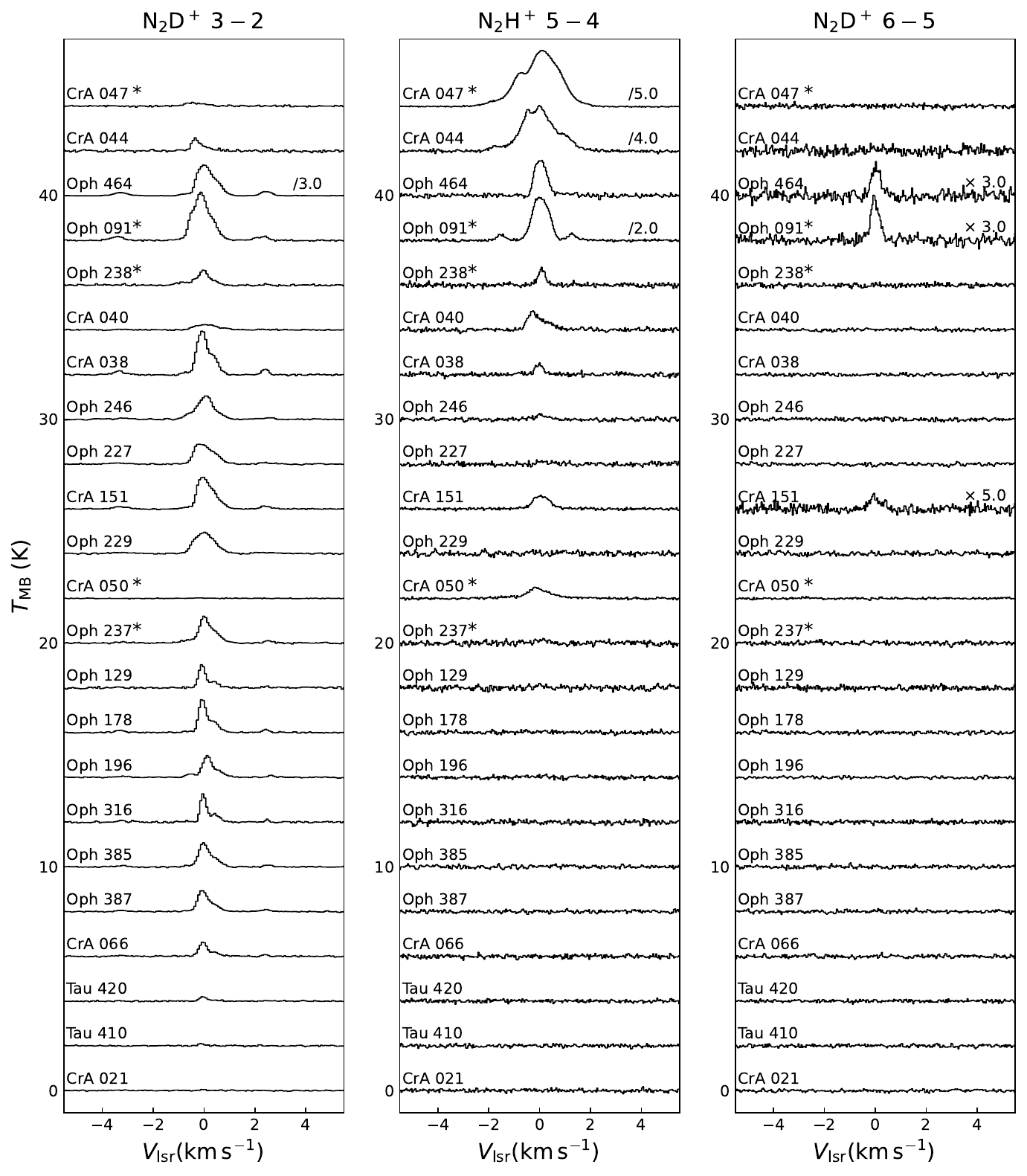} 
\end{center}
\caption{Spectra of \ntdp(3-2), \nthp(5-4), and \ntdp(6-5) toward a sub-sample of 23 pre-stellar core candidates. From top to bottom, the spectra follow the same order as in Fig.\,\ref{Fig:n2hp32}. The spectra have also been displaced in intensity by 2\,K and centered at 0 LSR velocity, to allow comparison. Note that some spectra have been divided or multiplied by factors of 2 to 5 (see labels) for clarity.}
\label{Fig:otherlines}
\end{figure*}

The other lines, \nthp(5-4), \ntdp(3-2) and \ntdp(6-5), observed in a sub-sample of 23 objects are displayed in Fig.\,\ref{Fig:otherlines}. 


\subsection{Constant excitation temperature analysis} \label{Subsec:CTex}
\subsubsection{Column density and deuterium fraction}
\noindent
The centrally concentrated density profiles of the pre-stellar core candidates in Table \ref{Tab:targets} require a proper non-LTE analysis. This will be done in the individual papers such as those from \cite{Redaelli2025} and \cite{Spezzano2025}, as a detailed physical structure of each core is needed. Here we consider a simple constant excitation temperature analysis to provide \nthp and \ntdp column densities, based on \nthp(3-2) and \ntdp(4-3), respectively, and the D/H ratio in all the objects showing \ntdp(4-3). As already mentioned, these two lines have very similar critical densities and have been observed with almost identical angular resolutions (Table \ref{Tab:lines}), providing a reliable measurement of the deuterium fraction in the core inner regions with densities above $\simeq$10$^5$\,\percc.

Gaussian fits taking into account the hyperfine structure (hfs) of the two lines have been carried out using the CLASS/GILDAS software\footnote{\url{https://www.iram.fr/IRAMFR/GILDAS/doc/html/class-html/class.html}} and the results are reported in Tables \ref{tab:multicolumn1} and \ref{tab:multicolumn2}. The objects in these tables have been ordered as those in Figures \ref{Fig:n2hp32} and \ref{Fig:n2dp43}, i.e., in order of decreasing \nthp(3-2) integrated intensity. The tables present the hfs fit output (columns 2 - 5: $J(T_{\rm MB})\times \tau$, where $J(T_{\rm MB})$ is the equivalent Rayleigh-Jeans main beam temperatures, and $\tau$ is the total optical depth, i.e., the sum of the optical depth of all the hyperfines, the centroid velocity, $\rm v_{LSR}$, the intrinsic line width, FWHM, and the total optical depth), the excitation temperature $T_{\rm ex}$ (column 6), derived from the radiative transfer equation and the information from columns 2 and 5, the total column density $\rm N$ (column 7). If the line is optically thin, or if the error on $\tau$ is larger than $\tau/3$, the excitation temperature must be assumed. Column 8 of Tables \ref{tab:multicolumn1} and \ref{tab:multicolumn2} shows the column density using the assumed $T_{\rm ex}$. In general, if $T_{\rm ex}$ can be measured for the \nthp(3-2) line but not for the \ntdp(4-3) line, we used the \nthp(3-2) excitation temperature for the calculation of the \ntdp column density. If $T_{\rm ex}$ cannot be measured at all, then we adopt for both lines $T_{\rm ex}$ = 5\,K, the typical value for similarly faint objects. Column 9 shows the D/H ratio ($R_{\rm D}$) calculated by simply taking the ratio of the \ntdp and \nthp column densities. $R_{\rm D}$ values range from about 0.016$\pm$0.001 (Oph\,082, see Table\ref{tab:multicolumn1}) to 4$\pm$2 (CrA\,151). 

\subsubsection{Correlations}
\begin{figure}
\begin{center}
\includegraphics[width=8cm]{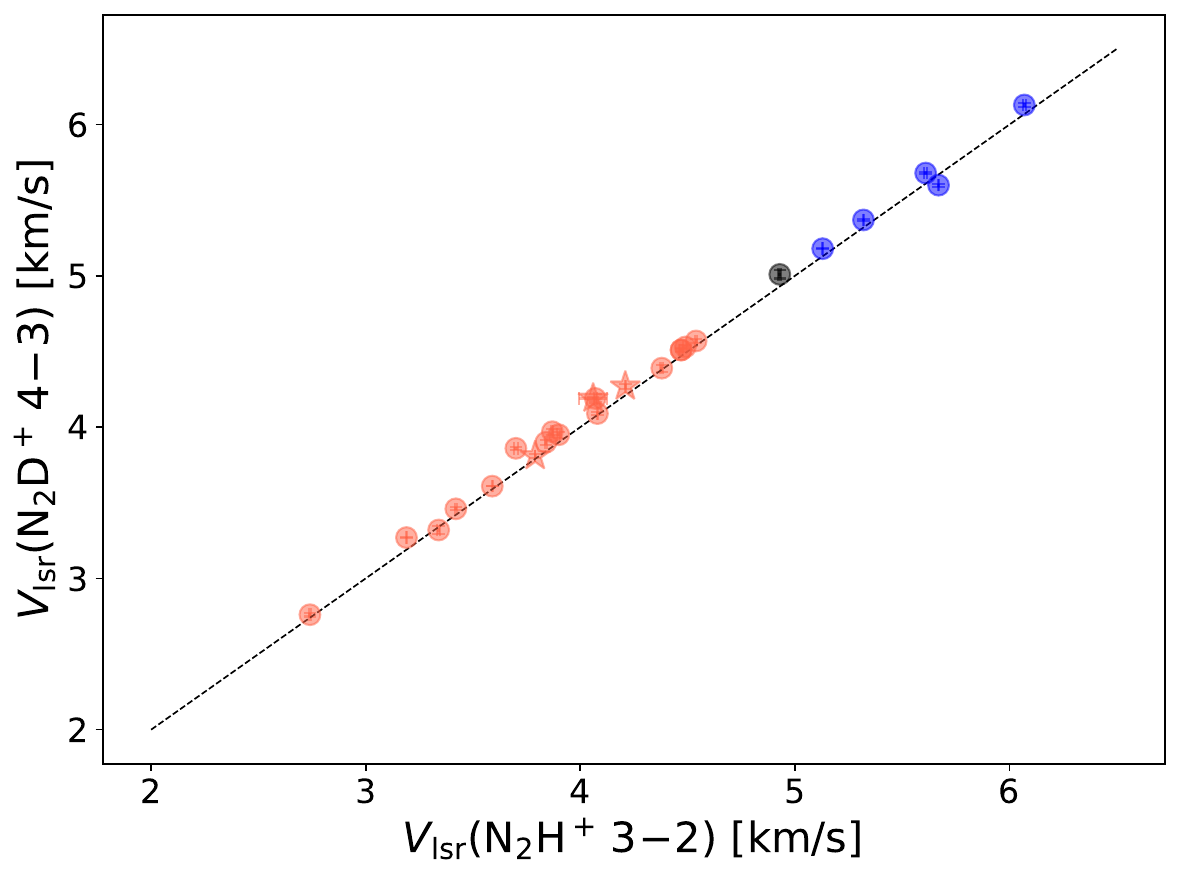} 
\includegraphics[width=8cm]{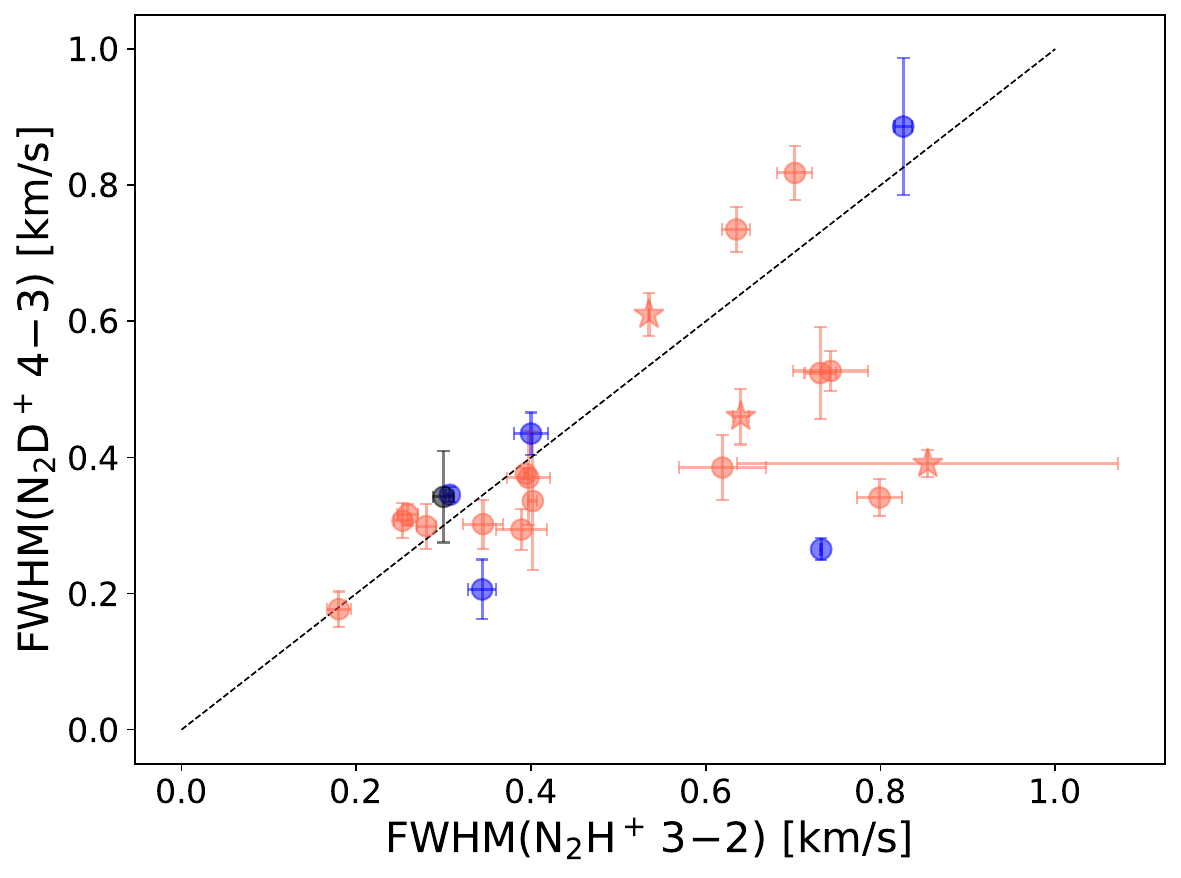} 
\end{center}
\caption{Comparison of centroid velocities ($\rm v_{LSR}$) and line widths (FWHM) of the \ntdp(4-3) and \nthp(3-2) lines. The different colors represent different molecular cloud complexes: Corona Australis (blue), Ophiuchus (red), Lupus (black). The stars show the cores associated with young stellar objects. The 1:1 correlation line is shown in dashed black.}
\label{Fig:vlsr-fwhm}
\end{figure}

\noindent
Figure \ref{Fig:vlsr-fwhm} compares the centroid velocities ($\rm v_{LSR}$) and the line widths (FWHM) of \ntdp(4-3) and \nthp(3-2), showing a good correlation, especially in $\rm v_{LSR}$. This suggests that these two lines trace similar gas. The larger FWHM found in \nthp(3-2) for some of the objects could be due to the larger volume occupied by \nthp, thus tracing the more turbulent motions or accretion flows in the outer regions of the core. However, higher angular resolution observations and mapping are needed to shed light on this difference. The cores associated with very young stellar objects do not stand out in these figures, once again showing that the physical properties of the original pre-stellar core have not been significantly modified. 

\begin{figure}
\begin{center}
\includegraphics[width=9cm]{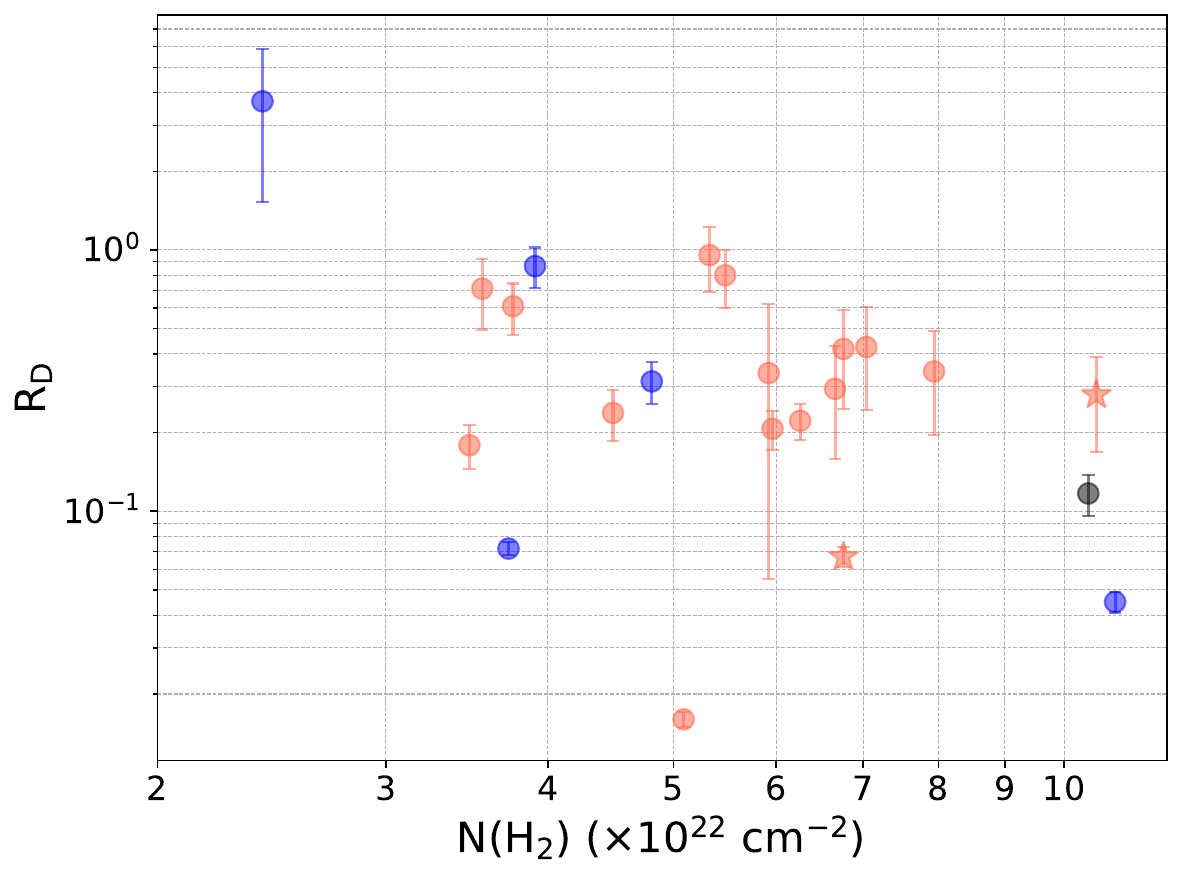} 
\end{center}
\caption{\nthp deuterium fraction ($R_{\rm D} \equiv$ N(\ntdp)/N(\nthp) column density ratio) obtained from the \ntdp(4-3) and \nthp(3-2) lines using constant excitation temperature analysis, as a function of H$_2$ column density, N(H$_2$), from {\em{Herschel}} data. The different colors represent different molecular cloud complexes: Corona Australis (blue), Ophiuchus (red), Lupus (black). The stars show the cores associated with young stellar objects.}
\label{Fig:LTE-RD}
\end{figure}

Figure \ref{Fig:LTE-RD} shows the \nthp deuterium fraction ($R_{\rm D} \equiv$ N(\ntdp)/N(\nthp)) as a function of the H$_2$ column density, N(H$_2$), obtained from {\em{Herschel}} data (see Table \ref{Tab:targets}). The values range from $\simeq$0.02 for core Oph\,082 to $\simeq$4 for core CrA\,151, with a mean of 0.49$\pm$0.09. However, there is no obvious correlation between the two quantities, although higher column densities should imply higher volume densities, thus larger deuterium fractions at the low dust temperatures reported in Table\,\ref{Tab:targets}. It is important to note that these values (derived from a constant excitation temperature analysis) are highly uncertain as non-LTE effects play a crucial role in the correct measurement of column densities in pre-stellar cores. This is clearly shown in \cite{Redaelli2025} and \cite{Spezzano2025}. For example, CrA\,151 has been studied in detail by \cite{Redaelli2025}, who found R$_{\rm D}$ $\simeq$ 0.5 (instead of our 4$\pm$2, see Table\,\ref{tab:multicolumn1}), while in Oph\,464 \cite{Spezzano2025} found R$_{\rm D}$ $\simeq$ 0.4 (instead of 0.24$\pm$0.03, Table\,\ref{tab:multicolumn1}). A detailed discussion on the discrepancies between the two methods to deduce column densities and D-fractions can be found in the two papers. The missing trend between $R_{\rm D}$ and N(H$_2$) in Fig.\,\ref{Fig:LTE-RD} may then be due to the assumption of constant excitation temperature and results are expected to change once non-LTE analysis will be carried out for all cores. Figure \ref{Fig:LTE-RD} is however still useful for our overview study and for comparison with previous work (see next section).   

\begin{table*}
\caption{Hyperfine fit results for the cores in Table\,\ref{Tab:targets}, ordered as in Figure\,\ref{Fig:n2hp32} (see text for details). Note that the \nthp(3-2) spectrum towards CrA\,047 has two velocity components.}
\label{tab:multicolumn1}
\scalebox{0.9}{
\begin{tabular}{ccccccccc}
    \hline
 Line&$J$(T$_{MB})$$\times$$\tau$&v$_{LSR}$ & $FWHM$& $\tau$& T$_{ex}$ (hfs) &N(T$_{ex}$ hfs)&N(T$_{ex}$)&R$_D$\\
 &K&km/s&km/s&&K&$\times$10$^{13}$ cm$^{-2}$&$\times$10$^{13}$ cm$^{-2}$&-\\
 \hline
 \multicolumn{9}{c}{CrA\,047*}\\
 N$_2$H$^+$ 3-2  & $50.400 \pm 0.047$  & $6.100 \pm 0.065$  & $1.440 \pm 0.218$  & $7.390 \pm 0.100$ & $12.44\pm 0.10$& $6.89\pm 1.05$ & - & - \\ 
N$_2$H$^+$ 3-2  & $16.300 \pm 0.047$  & $6.320 \pm 0.065$  & $0.534 \pm 0.218$  & $2.280 \pm 0.100$ & $12.81\pm 0.34$& $0.80\pm 0.33$ &- &- \\ 
\multicolumn{9}{c}{CrA\,044}\\
N$_2$H$^+$ 3-2  & $137.000 \pm 0.152$  & $5.670 \pm 0.001$  & $0.732 \pm 0.001$  & $18.400 \pm 0.022$ & $13.13\pm 0.01$& $8.94\pm 0.02$ &- &0.31(6) \\ 
N$_2$D$^+$ 4-3  & $1.460 \pm 0.069$  & $5.600 \pm 0.011$  & $0.265 \pm 0.016$  & $2.030 \pm 0.201$ & $4.95\pm 0.16$& $2.81\pm 0.51$ &-&-\\
\multicolumn{9}{c}{Oph\,087}\\
N$_2$H$^+$ 3-2  & $43.100 \pm 0.081$  & $3.130 \pm 0.000$  & $0.303 \pm 0.001$  & $7.750 \pm 0.018$ & $11.04\pm 0.02$& $1.46\pm 0.01$ &- & -  \\ 
\multicolumn{9}{c}{Oph\,464}\\
N$_2$H$^+$ 3-2  & $87.500 \pm 0.268$  & $3.590 \pm 0.002$  & $0.394 \pm 0.002$  & $24.400 \pm 0.088$ & $8.74\pm 0.02$& $5.96\pm 0.04$ &- & 0.24(3) \\ 
N$_2$D$^+$ 4-3  & $9.940 \pm 0.572$  & $3.610 \pm 0.002$  & $0.376 \pm 0.008$  & $2.780 \pm 0.225$ & $9.12\pm 0.44$& $1.45\pm 0.20$ &- & - \\ 
\multicolumn{9}{c}{Oph\,091*}\\
N$_2$H$^+$ 3-2  & $62.700 \pm 0.033$  & $3.790 \pm 0.000$  & $0.535 \pm 0.000$  & $15.600 \pm 0.005$ & $9.26\pm 0.00$& $5.13\pm 0.00$ &- & 0.0663(2)  \\ 
N$_2$D$^+$ 4-3  & $1.370 \pm 0.059$  & $3.810 \pm 0.013$  & $0.610 \pm 0.031$  & $0.1$  & -& -& $0.340(1)$ & - \\ 
\multicolumn{9}{c}{Oph\,238*}\\
N$_2$H$^+$ 3-2  & $6.900 \pm 0.152$  & $4.210 \pm 0.002$  & $0.640 \pm 0.009$  & $1.120 \pm 0.060$ & $11.71\pm 0.40$& $0.45\pm 0.03$ &-&0.067(6) \\ 
N$_2$D$^+$ 4-3  & $0.356 \pm 0.025$  & $4.270 \pm 0.015$  & $0.460 \pm 0.041$  & $0.1$  & -& -& $0.030(2)$ &-\\ 
\multicolumn{9}{c}{Oph\,082}\\
N$_2$H$^+$ 3-2  & $16.800 \pm 0.436$  & $3.190 \pm 0.002$  & $0.402 \pm 0.005$  & $5.120 \pm 0.157$ & $8.37\pm 0.16$& $1.29\pm 0.06$ &- & 0.016(1) \\ 
N$_2$D$^+$ 4-3  & $0.108 \pm 0.029$  & $3.270 \pm 0.048$  & $0.336 \pm 0.101$  & $0.1$  & -& -& $0.020(1)$ &- \\ 
\multicolumn{9}{c}{CrA\,040}\\
N$_2$H$^+$ 3-2  & $11.400 \pm 0.363$  & $5.130 \pm 0.003$  & $0.826 \pm 0.010$  & $3.810 \pm 0.152$ & $8.01\pm 0.19$& $2.00\pm 0.11$ &-& 0.045(4)  \\ 
N$_2$D$^+$ 4-3  & $0.160 \pm 0.017$  & $5.180 \pm 0.050$  & $0.886 \pm 0.101$  & $0.1$  & -& -& $0.090(5)$ &-  \\ 
\multicolumn{9}{c}{CrA\,038}\\
N$_2$H$^+$ 3-2  & $67.600 \pm 1.660$  & $5.320 \pm 0.002$  & $0.307 \pm 0.004$  & $29.900 \pm 0.788$ & $7.06\pm 0.11$& $6.24\pm 0.25$ &- & 0.072(4)  \\ 
N$_2$D$^+$ 4-3  & $1.240 \pm 0.033$  & $5.370 \pm 0.004$  & $0.345 \pm 0.011$  & $0.1$  & -& -& $0.45(2)$ &- \\ 
\multicolumn{9}{c}{Oph\,246}\\
N$_2$H$^+$ 3-2  & $5.380 \pm 0.875$  & $4.080 \pm 0.004$  & $0.743 \pm 0.043$  & $2.150 \pm 0.592$ & $7.38\pm 1.05$& $1.06\pm 0.39$ &- &0.4(2)  \\ 
N$_2$D$^+$ 4-3  & $0.681 \pm 0.031$  & $4.090 \pm 0.012$  & $0.527 \pm 0.029$  & $0.1$  & -& -& $0.4(1)$ & - \\ 
\multicolumn{9}{c}{Oph\,227}\\
N$_2$H$^+$ 3-2  & $6.210 \pm 0.371$  & $3.840 \pm 0.004$  & $0.635 \pm 0.016$  & $3.040 \pm 0.250$ & $6.77\pm 0.29$& $1.35\pm 0.16$ & - &0.21(3) \\ 
N$_2$D$^+$ 4-3  & $0.362 \pm 0.015$  & $3.900 \pm 0.016$  & $0.735 \pm 0.033$  & $0.1$  & -& -& $0.28(3)$ &- \\ 
\multicolumn{9}{c}{CrA\,151}\\
N$_2$H$^+$ 3-2  & $7.380 \pm 0.614$  & $5.610 \pm 0.006$  & $0.400 \pm 0.019$  & $4.290 \pm 0.432$ & $6.31\pm 0.32$& $1.28\pm 0.20$ &- & 4(2) \\ 
N$_2$D$^+$ 4-3  & $2.020 \pm 0.405$  & $5.680 \pm 0.006$  & $0.435 \pm 0.031$  & $2.420 \pm 0.738$ & $5.18\pm 0.58$& $4.74\pm 2.68$ & - & - \\ 
\multicolumn{9}{c}{Oph\,229}\\
N$_2$H$^+$ 3-2  & $5.210 \pm 0.384$  & $3.870 \pm 0.004$  & $0.702 \pm 0.020$  & $2.490 \pm 0.283$ & $6.83\pm 0.39$& $1.22\pm 0.19$ & - &0.24(5) \\ 
N$_2$D$^+$ 4-3  & $0.332 \pm 0.015$  & $3.970 \pm 0.020$  & $0.818 \pm 0.041$  & $0.1$  & -& -& $0.29(5)$ &- \\ 
\multicolumn{9}{c}{CrA\,050*}\\
N$_2$H$^+$ 3-2  & $3.090 \pm 0.258$  & $4.820 \pm 0.005$  & $0.791 \pm 0.027$  & $0.863 \pm 0.235$ & $8.73\pm 1.24$& $0.42\pm 0.14$ & - & - \\ 
\multicolumn{9}{c}{Oph\,237*}\\
N$_2$H$^+$ 3-2  & $2.990 \pm 0.041$  & $4.060 \pm 0.065$  & $0.854 \pm 0.218$  & $1.100 \pm 0.100$ & $7.66\pm 0.32$& $0.61\pm 0.17$ & - & 0.3(1) \\ 
N$_2$D$^+$ 4-3  & $0.711 \pm 0.029$  & $4.190 \pm 0.007$  & $0.391 \pm 0.020$  & $0.1$  & -& -& $0.17(5)$ & - \\
\multicolumn{9}{c}{Oph\,129}\\
N$_2$H$^+$ 3-2  & $6.380 \pm 0.486$  & $2.740 \pm 0.004$  & $0.253 \pm 0.010$  & $3.480 \pm 0.393$ & $6.47\pm 0.35$& $0.64\pm 0.10$ &-& 0.6(1) \\ 
N$_2$D$^+$ 4-3  & $0.558 \pm 0.041$  & $2.760 \pm 0.012$  & $0.307 \pm 0.026$  & $0.1$  & -& -& $0.39(6)$ &-  \\ 
\multicolumn{9}{c}{Oph\,178}\\
N$_2$H$^+$ 3-2  & $4.880 \pm 0.446$  & $4.470 \pm 0.005$  & $0.259 \pm 0.012$  & $3.240 \pm 0.441$ & $6.00\pm 0.37$& $0.66\pm 0.13$ &-&0.9(3)\\ 
N$_2$D$^+$ 4-3  & $0.883 \pm 0.036$  & $4.510 \pm 0.006$  & $0.316 \pm 0.016$  & $0.1$  & -& -& $0.6(1)$ & - \\ 
\multicolumn{9}{c}{Oph\,410}\\
N$_2$H$^+$ 3-2  & $7.770 \pm 0.566$  & $4.540 \pm 0.005$  & $0.280 \pm 0.011$  & $5.810 \pm 0.515$ & $5.74\pm 0.24$& $1.34\pm 0.18$ &- & 0.18(3)\\ 
N$_2$D$^+$ 4-3  & $0.361 \pm 0.035$  & $4.570 \pm 0.014$  & $0.299 \pm 0.033$  & $0.1$  & -& -& $0.24(3)$ & - \\ 
\multicolumn{9}{c}{Oph\,412}\\
N$_2$H$^+$ 3-2  & $3.770 \pm 0.510$  & $4.380 \pm 0.006$  & $0.397 \pm 0.025$  & $2.270 \pm 0.522$ & $6.23\pm 0.64$& $0.68\pm 0.22$ & - & 0.3(1) \\ 
N$_2$D$^+$ 4-3  & $0.238 \pm 0.035$  & $4.390 \pm 0.023$  & $0.370 \pm 0.069$  & $0.1$  & -& -& $0.20(7)$ & - \\ 
\multicolumn{9}{c}{Oph\,196}\\
N$_2$H$^+$ 3-2  & $2.020 \pm 0.183$  & $3.700 \pm 0.010$  & $0.799 \pm 0.026$  & $1.290 \pm 0.286$ & $6.09\pm 0.55$& $0.80\pm 0.24$ &- & 0.4(2) \\ 
N$_2$D$^+$ 4-3  & $0.441 \pm 0.030$  & $3.860 \pm 0.011$  & $0.341 \pm 0.027$  & $0.1$  & -& -& $0.3(1)$ & - \\ 
\hline
\end{tabular}
}
\end{table*}

\newpage

\begin{table*}
\caption{Continuation of Table\,\ref{tab:multicolumn1}.}
\label{tab:multicolumn2}
\scalebox{0.9}{
\begin{tabular}{ccccccccc}
    \hline
 Line&$J$(T$_{MB})$$\times$$\tau$&v$_{LSR}$ & $FWHM$& $\tau$& T$_{ex}$ (hfs) &N(T$_{ex}$ hfs)&N(T$_{ex}$)&R$_D$\\
 &K&km/s&km/s&&K&$\times$10$^{13}$ cm$^{-2}$&$\times$10$^{13}$ cm$^{-2}$&-\\
 \hline
\multicolumn{9}{c}{Lup\,288}\\
N$_2$H$^+$ 3-2  & $10.100 \pm 0.735$  & $4.930 \pm 0.006$  & $0.300 \pm 0.012$  & $10.800 \pm 0.863$ & $5.09\pm 0.18$& $3.17\pm 0.40$ & - & 0.12(2) \\ 
N$_2$D$^+$ 4-3  & $0.160 \pm 0.026$  & $5.010 \pm 0.027$  & $0.342 \pm 0.067$  & $0.1$  & -& -& $0.37(5)$ & - \\ 
\multicolumn{9}{c}{Oph\,316}\\
N$_2$H$^+$ 3-2  & $4.650 \pm 0.658$  & $3.420 \pm 0.005$  & $0.180 \pm 0.014$  & $3.620 \pm 0.749$ & $5.66\pm 0.50$& $0.55\pm 0.16$ & -& 0.4(2) \\ 
N$_2$D$^+$ 4-3  & $0.582 \pm 0.075$  & $3.460 \pm 0.010$  & $0.177 \pm 0.026$  & $0.1$  & -& -& $0.23(7)$ &- \\ 
\multicolumn{9}{c}{Oph\,215}\\
N$_2$H$^+$ 3-2  & $1.470 \pm 0.033$  & $4.070 \pm 0.007$  & $0.731 \pm 0.018$  & $0.194 \pm 0.013$ & $13.27\pm 0.56$& $0.09\pm 0.01$ & - & 0.22(3)\\ 
N$_2$D$^+$ 4-3  & $0.213 \pm 0.022$  & $4.190 \pm 0.027$  & $0.524 \pm 0.067$  & $0.1$  & -& -& $0.020(2)$ & -\\ 
\multicolumn{9}{c}{Oph\,385}\\
N$_2$H$^+$ 3-2  & $3.680 \pm 0.537$  & $4.470 \pm 0.008$  & $0.389 \pm 0.029$  & $3.030 \pm 0.640$ & $5.55\pm 0.50$& $1.02\pm 0.31$ &- & 0.3(1)\\ 
N$_2$D$^+$ 4-3  & $0.526 \pm 0.044$  & $4.510 \pm 0.011$  & $0.294 \pm 0.030$  & $0.1$  & -& -& $0.4(1)$ & - \\ 
\multicolumn{9}{c}{Oph\,387}\\
N$_2$H$^+$ 3-2  & $5.200 \pm 0.603$  & $4.490 \pm 0.009$  & $0.345 \pm 0.023$  & $4.980 \pm 0.706$ & $5.27\pm 0.32$& $1.59\pm 0.34$ &-& 0.7(2) \\ 
N$_2$D$^+$ 4-3  & $0.551 \pm 0.055$  & $4.530 \pm 0.014$  & $0.302 \pm 0.036$  & $0.1$  & -& -& $1.1(2)$ &- \\ 
\multicolumn{9}{c}{Oph\,201}\\
N$_2$H$^+$ 3-2  & $2.210 \pm 0.428$  & $3.900 \pm 0.008$  & $0.619 \pm 0.050$  & $1.460 \pm 0.641$ & $6.01\pm 1.08$& $0.71\pm 0.42$ &- & 0.3(3) \\ 
N$_2$D$^+$ 4-3  & $0.281 \pm 0.029$  & $3.950 \pm 0.020$  & $0.385 \pm 0.048$  & $0.1$  & -& -& $0.2(1)$ & -\\
\multicolumn{9}{c}{Oph\,455}\\
N$_2$H$^+$ 3-2  & $1.830 \pm 0.212$  & $3.710 \pm 0.011$  & $0.911 \pm 0.043$  & $1.340 \pm 0.346$ & $5.79\pm 0.59$& $1.00\pm 0.35$ & - &-\\ 
\multicolumn{9}{c}{Oph\,319}\\
N$_2$H$^+$ 3-2  & $1.650 \pm 0.091$  & $3.340 \pm 0.006$  & $0.201 \pm 0.012$  & $0.1$  & -& -& $0.38(2)$ & 0.8(2) \\ 
N$_2$D$^+$ 4-3  & $0.213 \pm 0.042$  & $3.32 \pm 0.029$  & $0.2$  & $0.1$  & -& -& $0.29(6)$ & - \\ 
\multicolumn{9}{c}{Oph\,332}\\
N$_2$H$^+$ 3-2  & $3.490 \pm 0.502$  & $3.270 \pm 0.008$  & $0.234 \pm 0.017$  & $4.070 \pm 0.838$ & $4.95\pm 0.38$& $0.97\pm 0.29$ &-&- \\ 
\multicolumn{9}{c}{CrA\,066}\\
N$_2$H$^+$ 3-2  & $1.160 \pm 0.048$  & $6.070 \pm 0.007$  & $0.344 \pm 0.016$  & $0.1$  & -& -& $0.45(2)$ &0.9(2)  \\ 
N$_2$D$^+$ 4-3  & $0.277 \pm 0.048$  & $6.130 \pm 0.016$  & $0.206 \pm 0.044$  & $0.1$  & -& -& $0.39(7)$ &- \\ 
\multicolumn{9}{c}{Oph\,169*}\\
N$_2$H$^+$ 3-2  & $5.420 \pm 4.980$  & $3.920 \pm 0.010$  & $0.263 \pm 0.013$  & $8.890 \pm 0.493$ & $4.48\pm 1.13$& $2.87\pm 2.14$ & -& 0.2(1)  \\ 
N$_2$D$^+$ 4-3  & $0.279 \pm 0.046$  & $3.920 \pm 0.020$  & $0.243 \pm 0.048$  & $0.1$  & -& -& $0.46(8)$ & -  \\ 
\multicolumn{9}{c}{Oph\,219}\\
N$_2$H$^+$ 3-2  & $1.760 \pm 0.317$  & $4.680 \pm 0.012$  & $0.510 \pm 0.042$  & $2.640 \pm 0.743$ & $4.60\pm 0.43$& $1.57\pm 0.63$ &-& -\\ 
\multicolumn{9}{c}{Lup\,032}\\
N$_2$H$^+$ 3-2  & $1.630 \pm 0.353$  & $4.560 \pm 0.010$  & $0.346 \pm 0.037$  & $2.120 \pm 0.835$ & $4.79\pm 0.64$& $0.79\pm 0.44$ &- & -\\ 
\multicolumn{9}{c}{Oph\,485}\\
N$_2$H$^+$ 3-2  & $3.850 \pm 0.742$  & $2.510 \pm 0.016$  & $0.267 \pm 0.031$  & $6.830 \pm 1.470$ & $4.39\pm 0.34$& $2.33\pm 0.79$ &- & - \\ 
\multicolumn{9}{c}{Tau\,420}\\
N$_2$H$^+$ 3-2  & $2.220 \pm 0.561$  & $7.310 \pm 0.017$  & $0.277 \pm 0.040$  & $4.990 \pm 1.570$ & $4.13\pm 0.40$& $2.02\pm 0.94$ &-&- \\  
\multicolumn{9}{c}{Tau\,410}\\
N$_2$H$^+$ 3-2  & $1.090 \pm 0.485$  & $7.350 \pm 0.017$  & $0.216 \pm 0.045$  & $2.910 \pm 2.140$ & - & - &0.17(3)& - \\ 
\multicolumn{9}{c}{CrA\,021}\\
N$_2$H$^+$ 3-2  & $0.468 \pm 0.073$  & $6.110 \pm 0.016$  & $0.189 \pm 0.031$  & $0.1$  & -& -& $0.10(1)$ & -  \\ 
\hline
\end{tabular}
}
\end{table*}

\newpage
 
\section{Discussion} \label{Sec:discussion}

Using a similar (constant excitation temperature) analysis, \cite{Crapsi2005} defined as pre-stellar all dense starless cores with a deuterium fraction larger than 10\%. If we adopt the same criterium, and with the caveats described in the previous section, we see from Tables\,\ref{tab:multicolumn1} and \ref{tab:multicolumn2} and Fig.\,\ref{Fig:LTE-RD} that 20 out of the 26 cores with $R_{\rm D}$ values (77\% of the sample) can be defined as "pre-stellar". Among these, however, Oph\,237, Oph\,169 and maybe CrA\,151 \citep[see][]{Redaelli2025} are associated with a young stellar object. Therefore, we conclude that 17 of the studied objects are {\it bona fide} pre-stellar cores. This sample will be studied in detail with planned interferometric observations. The other starless cores with deuterium fractions less than 10\% are also interesting and deserve future scrutiny. The reason for this is that their high central densities derived from the {\em Herschel} data imply dynamically evolved structures either on the verge of protostar formation or hosting very young protostars, so the low deuterium fractions may point to different chemical evolution possibly linked to different environmental conditions. 

It is not surprising that six \citep[and possibly seven, if CrA\,151 will be confirmed protostellar; see][]{Redaelli2025} of the 40 {\em Herschel}-selected cores have been found associated with young stellar objects. The transition from pre-stellar to protostellar is expected to be fast, with very young protostellar objects still being surrounded by the cold and dense envelopes part of the original pre-stellar cores. In fact, four of the six cores associated with young stellar objects (Oph\,091, Oph\,169, Oph\,237, and Oph\,238), have $R_{\rm D}$ between 0.07 and 0.3, so they do not show systematically lower deuterium fraction values as expected in case of warm dense gas. This implies that the associated protostar is very young or with a very low luminosity and has not significantly modified the physical conditions of the surrounding dense envelope traced by \nthp (3-2) and \ntdp (4-3) with APEX\footnote{The APEX beam at these frequencies is about 20\arcsec (see Table\,\ref{Tab:lines}), or $\sim$3000\,au at the distance of the selected targets.}.  \cite{Crapsi2005} also included in their list of pre-stellar cores an object, L1521F \citep{Crapsi2004}, which was soon after found to harbor a very low luminosity object \citep[or VeLLO;][]{Bourke2005}. We stress that the D-fraction measured with \nthp, a molecule produced in the gas phase, is particularly sensitive to the local kinetic temperature 
($T_{\rm kin}$), as proton-deuteron exchange reactions such as H$_3^+$ + HD $\rightleftharpoons$ H$_2$D$^+$ + H$_2$ start to proceed from right to left when $T_{\rm kin}$ $\geq$ 30\,K \citep{Watson1973}, quickly dropping the abundance of \ntdp (formed from the reaction of N$_2$ with H$_3^+$ deuterated isotopologues). Only two cores in our sample associated with young stellar objects (Cra\,047 and CrA\,050) have no detection of \ntdp lines, thus suggesting the presence of a warm ($T_{\rm kin}$ $>$ 30\,K) envelope possibly due to central heating or dynamical/chemical evolution in different (warm?) environments where deuterium fractionation is not favored. This will be checked in future studies.



\begin{figure}
\begin{center}
\includegraphics[width=9cm]{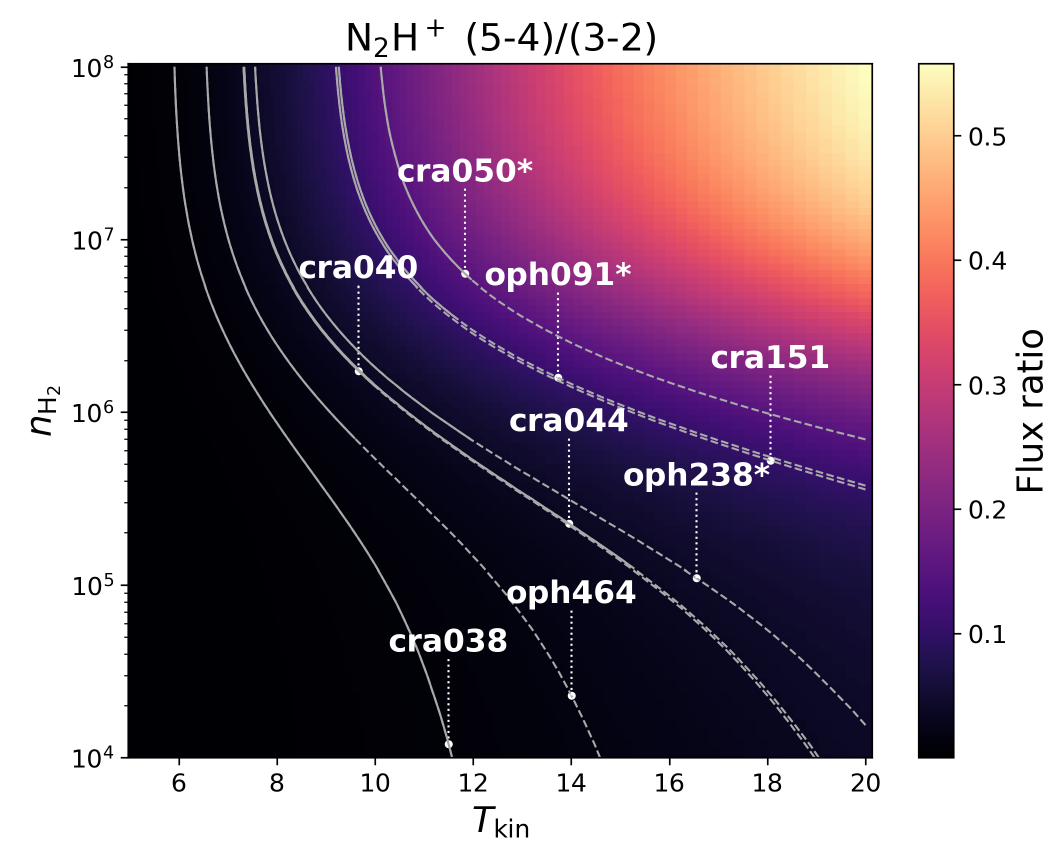} 
\end{center}
\caption{H${_2}$ volume density ($n_{H_2}$) versus kinetic temperature ($T_{\rm kin}$) traced by the \nthp (5-4)/(3-2) line flux ratio and based on the RADEX non-LTE code \citep{vdTak2007}. The objects where this flux ratio was possible to measure are indicated in the figure by the labeled gray curves. The curves are dashed if $T_{\rm kin} > T_{\rm dust}$ and full if $T_{\rm kin} < T_{\rm dust}$, where $T_{\rm dust}$ is the dust temperature measured with {\em Herschel} (see Table\,\ref{Tab:targets}).}

\label{Fig:Radex_nthp}
\end{figure}

\begin{figure*}
\begin{center}
\includegraphics[width=17cm]{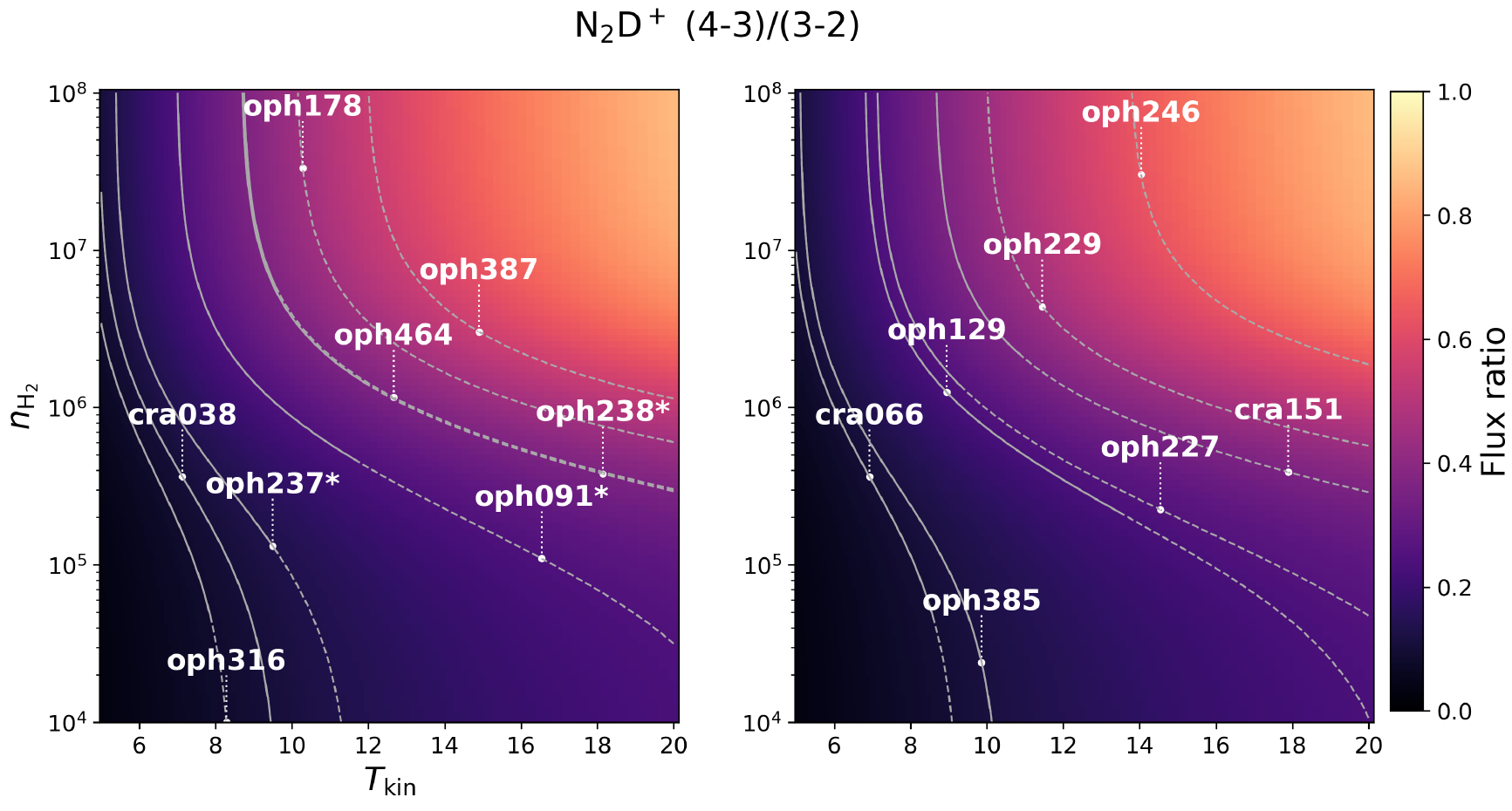} 
\end{center}
\caption{Same as Figure\,\ref{Fig:Radex_nthp} but for the \ntdp (4-3)/(3-2) line flux ratio.}
\label{Fig:Radex_ntdp}
\end{figure*}

The high central volume densities obtained from {\em Herschel} data (all above the threshold value of 3$\times$10$^5$\,\percc; see Section\,\ref{Subsec:sample} and Table\,\ref{derived_core_properties}), can be compared with the volume densities obtained using line ratios for those cores where multiple lines have been observed in \nthp and \ntdp . Figures\,\ref{Fig:Radex_nthp} and \ref{Fig:Radex_ntdp} present volume density ($n_{\rm H_2}$) vs. kinetic temperature ($T_{\rm kin}$) curves for the \nthp (5-4)/(3-2) and \ntdp (4-3)/(3-2) line flux ratios, respectively, obtained with the RADEX code \citep{vdTak2007}. The line flux ratios for each individual source are marked by labeled gray curves, which are full if $T_{\rm kin} < T_{\rm dust}$ and dashed if $T_{\rm kin} > T_{\rm dust}$, where $T_{\rm dust}$ is the dust temperature from {\em Herschel} (see Table\,\ref{Tab:targets}). It is interesting to note that some objects need a $T_{\rm kin}$ significantly lower than $T_{\rm dust}$ to have central densities above the selection threshold of 3$\times$10$^5$\,\percc (see Section\,\ref{Subsec:sample}). This is the case for CrA\,038 (Figs.\,\ref{Fig:Radex_nthp} and \ref{Fig:Radex_ntdp}), an object with a deuterium fraction below 0.1 (Table\,\ref{tab:multicolumn1}) and three {\it bona-fide} pre-stellar cores  (Cra\,066, Oph\,316, Oph\,385; Fig.\,\ref{Fig:Radex_ntdp}). Kinetic temperatures lower than dust temperatures are expected in case of external heating. For example, an enhanced interstellar radiation field could heat the dust in the outer core envelope, so that the {\em Herschel} dust temperature measurement toward the core center will be skewed to values at least a few K higher than the central temperature, as also found in other studies \citep[e.g.,][in the Ophiuchus Molecular Cloud]{Choudhury2021}. More detailed analysis will be carried out in future papers focused on these objects to investigate this point. In particular, interferometric observations of NH$_3$ will be required to measure the central gas temperature and more accurately reconstruct the temperature profiles across the cores, following procedures similar to those described in \cite{Crapsi2007} and \cite{Lin2023a}.

\section{Conclusions} \label{Sec:conclusion}

Pre-stellar cores represent the initial conditions in the process of star and planet formation, but they are difficult to find as they are short-lived. With the aim of increasing the number of known pre-stellar cores in nearby ($<$200\,pc) clouds and study their properties as a function of environment, we used archival {\em Herschel} data on starless cores from the Gould Belt Survey \citep{Andre2010} to identify objects with central densities higher than or equal to 3$\times$10$^5$\,\percc \citep[the density of the prototypical pre-stellar core L1544 within the central 20\arcsec; e.g.,][]{Caselli2019}. We found 40 cores (out of 1746), which have then been observed with APEX to enquire about their nature. This has been done by scrutinizing the line profile of high density tracers with high spectral resolution (61\,kHz) and high sensitivity (20-60\,mK) and by measuring the deuterium fraction using \nthp(3-2) and \ntdp(4-3) lines. Of the 40 dense cores,  17 can be considered "{\it bona-fide}" pre-stellar cores, as they have deuterium fractions larger than 10\% \citep[following the original definition of][]{Crapsi2005}. Six cores \citep[or seven if also CrA\,151 is included; see][]{Redaelli2025} of the 40 originally selected cores with {\em Herschel} host a very young (and/or very low luminosity) stellar object, that has not yet significantly modified the chemical composition of its envelope, as the corresponding deuterium fractions measured with APEX are relatively large (between 7 and 30\%), except for CrA\,047 and CrA\,050, not detected in \ntdp lines. 
The other pre-stellar cores (17 objects) with deuterium fractions less than 10\% require further investigation, to understand what causes the lower level of deuteration in such dynamically evolved structures. 

Here we have presented an overview of our findings, following a simple analysis. More detailed analysis on each individual core is in progress and the first results can be found in \cite{Spezzano2025} and \cite{Redaelli2025}. Interferometric observations, in particular with ALMA, similar to those described in \cite{Caselli1999,Caselli2022} are also planned. ALMA observations are crucial to study the structure of the central regions, the so-called "kernel" \citep{Caselli2019}, its connection to the rest of the core and its dynamical evolution toward the formation of the young stellar object and the protoplanetary disk. The "kernel" chemical and physical structure is required to put quantitative constraints on simulations of star and planet formation. 

This work demonstrates that a combination of dust continuum emission and high-density-tracer spectral line data offers a powerful tool to identify pre-stellar cores, a crucial step in the formation of stellar systems. Detailed studies of pre-stellar cores are needed not just for understanding the process of star and planet formation by setting the initial conditions, but also for quantifying the chemical inheritance from clouds to planets and our astrochemical origins. 

\begin{acknowledgement}
The authors acknowledge our referee, Jes  J{\o}rgensen, for careful reading of the paper and important suggestions. We also thank the Max Planck Society for the support. 
\end{acknowledgement}

\bibliographystyle{aa} 

\bibliography{bibliography.bib} 

\newpage

\begin{appendix}
\section{Core Physical Properties derived from the HGBS data} \label{Appendix}

Physical properties derived from the Herschel Gould Belt Survey (HGBS) data are listed in Table~\ref{derived_core_properties}. The cores associated with YSOs are listed in Table~\ref{targets_w_ysos}. This table gives the observed core radius from the HGBS catalogs, the angular distance to the nearest known YSO (provided that this is shorter than $70\arcsec$), and the number of YSOs within the core radius. The search for YSOs was performed using the SIMBAD Astronomical Database\footnote{https://simbad.cds.unistra.fr/simbad/}. A core was considered starless if no YSO is found within its radius. We also inspected the mid-infrared surface brightness maps of the 40 targets extracted from the {\sl Spitzer} and WISE\footnote{The Wide-field Infrared Survey Explorer} data archives available at {https://irsa.ipac.caltech.edu}. The starless cores, with the single exception of CrA\,044, did not show compact mid-infrared emission.

\begin{table}
{\centering
\caption[]{Core properties derived from Herschel Gould Belt Survey data.}
\label{derived_core_properties}
\begin{tabular}{lccrr}\hline
   \noalign{\smallskip}

HGBS   &   $R^a$ &  $M^a$ & \multicolumn{1}{c}{$T_{\rm dust}^a$} & 
\multicolumn{1}{c}{$n_{\htwo}^a$} \\
name   &   $(\arcsec)$ &  $(M_\odot)$  &  \multicolumn{1}{c}{(K)} &   $(10^5\,\percc)$ \\ \hline
\noalign{\smallskip}
Tau\,109 &  41.3 &  0.06 (0.02) &  11.0 (0.5) &  4.0 \\
Tau\,410 &  51.6 &  1.24 (0.65) &   6.8 (0.5) &  3.1 \\
Tau\,420 &  56.0 &  2.75 (0.78) &  10.1 (0.6) &  3.7 \\
 \noalign{\smallskip}                                                    
Lup\,288 &  39.5 &  2.50 (0.03) &  8.0  (0.1)  &12.0 \\
Lup\,032 &  31.3 &  2.20 (0.02) &  7.6  (0.1)  & 3.4 \\
Lup\,039 &  47.5 &  1.19 (0.15) &  7.1  (0.2)  & 4.8 \\
  \noalign{\smallskip}                                                  
Oph\,082 &  22.2 &  0.25 (0.12) &  14.5 (4.5) &  6.0 \\
Oph\,087 &  25.2 &  1.16 (0.21) &  11.9 (0.6) &  5.8 \\
Oph\,091 &  25.2 &  7.87 (1.06) &  12.7 (0.6) & 42.3 \\
Oph\,129 &  28.2 &  0.49 (0.04) &  13.6 (0.3) &  4.1 \\
Oph\,146 &  25.2 &  0.97 (0.30) &   7.0 (0.4) &  4.6 \\
Oph\,169 &  53.3 &  2.96 (0.40) &  11.6 (0.4) &  5.7 \\
Oph\,178 &  26.7 &  0.62 (0.12) &   9.7 (0.5) &  5.9 \\
Oph\,196 &  29.6 &  0.66 (0.33) &  11.5 (4.5) &  7.4 \\
Oph\,201 &  20.7 &  1.40 (0.18) &   7.9 (0.3) &  7.1 \\
Oph\,215 &  25.2 &  1.13 (0.25) &   8.8 (0.3) &  7.0 \\
Oph\,219 &  17.8 &  0.14 (0.07) &  14.5 (4.5) &  3.2 \\
Oph\,227 &  25.2 &  1.15 (0.17) &   9.4 (0.4) &  6.6 \\
Oph\,229 &  31.1 &  1.55 (0.29) &   9.5 (0.4) &  4.6 \\
Oph\,237 &  28.2 &  1.33 (0.22) &   9.7 (0.4) & 11.4 \\
Oph\,238 &  25.2 &  1.17 (0.06) &  12.0 (0.1) &  7.6 \\
Oph\,246 &  28.2 &  2.82 (0.30) &   8.4 (0.2) & 17.3 \\
Oph\,316 &  41.5 &  3.06 (0.46) &   7.9 (0.2) &  5.9 \\
Oph\,319 &  50.4 &  3.13 (0.60) &   7.6 (0.3) &  4.2 \\
Oph\,332 &  48.9 &  2.19 (0.34) &   8.9 (0.3) &  3.9 \\
Oph\,385 &  22.2 &  0.46 (0.23) &  11.5 (4.5) &  9.4 \\
Oph\,387 &  20.7 &  0.91 (0.46) &  11.5 (4.5) &  4.2 \\
Oph\,410 &  23.7 &  0.51 (0.10) &   9.9 (0.4) &  3.9 \\
Oph\,412 &  29.6 &  1.12 (0.22) &   9.0 (0.4) &  6.9 \\
Oph\,455 &  26.7 &  4.51 (1.99) &   6.4 (0.4) & 11.7 \\
Oph\,464 &  26.7 &  3.97 (0.32) &   9.6 (0.2) & 25.1 \\
Oph\,485 &  51.9 &  1.08 (0.54) &  11.5 (4.5) &  3.7 \\
\noalign{\smallskip} 
CrA\,021 &  19.0 &  0.08 (0.04) &  13.4 (1.0) &  8.3 \\ 
CrA\,038 &  25.4 &  0.67 (0.33) &  13.4 (1.0) &  7.2 \\
CrA\,040 &  22.2 &  1.34 (0.58) &  10.1 (1.1) & 23.2 \\
CrA\,044 &  19.0 &  0.66 (0.14) &  15.3 (0.9) & 11.9 \\
CrA\,047 &  20.6 &  0.98 (0.19) &  17.2 (1.0) & 16.7 \\
CrA\,050 &  17.5 &  0.62 (0.28) &  12.0 (1.4) & 11.4 \\
CrA\,066 &  33.3 &  0.69 (0.22) &   8.7 (0.5) &  5.7 \\
CrA\,151 &  20.6 &  0.41 (0.08) &  10.9 (0.4) &  5.8 \\ \hline
\noalign{\smallskip}
\end{tabular}

}

$^a$ The core radii ($R$), masses ($M$), dust temperatures ($T_{\rm dust}$), and beam-averaged peak densities ($n_{\htwo}$, in units of $10^5\,\percc$) are taken from 
the catalogues of \cite{2016MNRAS.459..342M} (Taurus), \cite{2018A&A...619A..52B} (Lupus), \cite{2020A&A...638A..74L} (Ophiuchus), and \cite{2018A&A...615A.125B} (Corona Australis).

\end{table}

\begin{table}
{\centering
\caption[]{Targets associated with YSOs.}
\label{targets_w_ysos}
\begin{tabular}{lrcl} \hline
   \noalign{\smallskip}
HGBS & $R_{\rm obs}$ & YSOs$^{a}$ & nearest YSO \\  \hline 
   \noalign{\smallskip}
Tau 420 & $56\arcsec$ &  0  & $68\arcsec$, TT J04184023+2824245 \\  
   \noalign{\smallskip}
Oph 082 & $22\arcsec$ & 0   & $55\arcsec$,  J162623.4-242100  \\    
Oph 087 & $25\arcsec$ &  0  & $38\arcsec$,  J162625.4-242301       \\    
Oph 091 & $25\arcsec$ &  2  & $2.5\arcsec$,  GDS J162627.8-242359      \\    
Oph 129 & $28\arcsec$ & 0   & $43\arcsec$,  TT GSS 38      \\    
Oph 146 & $25\arcsec$ &  0  & $59\arcsec$,  J162648.4-242838  \\    
Oph 169 & $53\arcsec$ &  3  & $37\arcsec$, TT WL22    \\    
Oph 178 & $27\arcsec$ &  0  & $56\arcsec$,  J162705.6-244013     \\    
Oph 196 & $30\arcsec$ &  0  & $39\arcsec$,  BBRCG 25       \\    
Oph 201 & $21\arcsec$ &  0  & $30\arcsec$,  J162715.5-243053      \\    
Oph 215 & $25\arcsec$ &  0  & $27\arcsec$, J162721.8-242727      \\    
Oph 219 & $18\arcsec$ &  0  & $57\arcsec$, J162718.3-243914    \\    
Oph 227 & $25\arcsec$ & 0   & $40\arcsec$, [AMD2002] J162722-242712    \\    
Oph 229 & $31\arcsec$ & 0   & $39\arcsec$, [GPJ2008] Source 8       \\    
Oph 237 & $28\arcsec$ &  1  & $26\arcsec$, J162727.53-242611.2      \\    
Oph 238 & $25\arcsec$ &  2  & $3.4\arcsec$, [GPJ2008] Source 8      \\    
Oph 246 & $28\arcsec$ &  0  & $68\arcsec$, TT ISO-Oph 150 \\    
Oph 410 & $24\arcsec$ &  0  & $61\arcsec$, J163200.9-245642     \\    
Oph 455 & $27\arcsec$ &  0  & $51\arcsec$, IRAS 16293-2422B \\  
   \noalign{\smallskip}
CrA 038 & $25\arcsec$ &  0  & $44\arcsec$,  A 445     \\    
CrA 040 & $22\arcsec$ &  0  & $28\arcsec$,  A 403     \\    
CrA 044$^{b}$ & $19\arcsec$ &  0  & $27\arcsec$, J190155.76-365727.7  \\    
CrA 047 & $21\arcsec$ &  1  & $19\arcsec$, J190155.76-365727.7  \\    
CrA 050 & $18\arcsec$ &  1  & $4.8\arcsec$, SMM 2, CrA-43      \\  \hline
   \noalign{\smallskip}
\end{tabular}

}   

$^{a}$ Number of YSOs within the core radius given in column 2 of Table\,\ref{derived_core_properties}. 

$^{b}$ CrA\,044 is associated with an unidentified compact 8 and 24\,$\mu$m source detected with {\sl Spitzer}. The source resides $13\arcsec$ from the core centre on its western side. Future spectroscopic observations with higher angular resolution will be carried out to assess the association of the source with the core. 

\end{table}

\end{appendix}

\end{document}